\documentclass[11pt]{article}
\usepackage[pdftex]{graphicx}
\usepackage{booktabs}
\pdfoutput =1
\usepackage{array, amsmath, graphicx, mathtools, amssymb, mathtools, setspace, booktabs, float, caption}
\usepackage[title]{appendix}
\usepackage{palatino}
\usepackage[noblocks]{authblk}
\usepackage{subcaption, caption, abstract, txfonts}

\graphicspath{{Figures/}}

\usepackage{url}
 
\usepackage{mathpazo, wrapfig, subcaption, tikz} 
\usepackage[scaled]{helvet} 
\usepackage{courier} 
\normalfont
\usepackage[T1]{fontenc}
\usetikzlibrary{%
   arrows,%
   calc,%
   fit,%
   patterns,%
   plotmarks,%
   shapes.geometric,%
   shapes.misc,%
   shapes.symbols,%
   shapes.arrows,%
   shapes.callouts,%
   shapes.multipart,%
   shapes.gates.logic.US,%
   shapes.gates.logic.IEC,%
   er,%
   automata,%
   backgrounds,%
   chains,%
   topaths,%
   trees,%
   petri,%
   mindmap,%
   matrix,%
   calendar,%
   folding,%
   fadings,%
   through,%
   positioning,%
   scopes,%
   decorations.fractals,%
   decorations.shapes,%
   decorations.text,%
   decorations.pathmorphing,%
   decorations.pathreplacing,%
   decorations.footprints,%
   decorations.markings,%
   shadows} 

\xdefinecolor{mygreen}{rgb}{0, 0, 1}
\xdefinecolor{myyell}{rgb}{1, 1, 0}
\xdefinecolor{mygreen}{rgb}{0.2, 0.8, 0.2}
\xdefinecolor{LightGreen}{rgb}{0.7, 1, 0.7}
\xdefinecolor{LightBlue}{rgb}{0.6, 0.6, 1}
\xdefinecolor{LightRed}{rgb}{ 1, 0.7, 0.7}
\xdefinecolor{LightGrey}{rgb}{0.7, 0.7, 0.7}
\xdefinecolor{LightYellow}{rgb}{1, 1, 0.7}
\xdefinecolor{Lav}{rgb}{.9, .9, 0.98}
\xdefinecolor{DarkV}{rgb}{.6,0,.8}

\xdefinecolor{DarkBlue}{rgb}{0,0,.45}

\newtheorem{proposition}{Proposition}
\newtheorem{lemma}{Lemma}
\newtheorem{example}{Example}

\newcommand{\W}{\hphantom{0}}
\providecommand{\keywords}[1]{\textbf{\textit{Keywords:}} #1}

 \newcommand{\indep}{{\bot\negthickspace\negthickspace\bot}}

\newcommand{\id}[1]{I_{#1}}

\newcommand{\Id}{$I_\text{dep}$ }
\newcommand{\Imm}{$I_\text{mmi}$ }

\newcommand{\red}{\text{red} }
\newcommand{\unqA}{\text{unq0}}
\newcommand{\unqB}{\text{unq1}}
\newcommand{\syn}{\text{syn}}

\begin{document}
\title{Exact partial information decompositions for Gaussian systems based on dependency constraints }
\author[1]{J. W. Kay}
\author[2]{R. A. A. Ince}
\affil[1]{Department of Statistics, University of Glasgow, UK}
\affil[2]{Institute of Neuroscience and Psychology, University of Glasgow, UK}

\maketitle
\begin{abstract}
The Partial Information Decomposition (PID)~\cite{WB} provides a theoretical framework to characterize and quantify the structure of multivariate information sharing. 
A new method ($I_{\text{dep}}$) has recently been proposed for computing a two-predictor partial information decomposition (PID) over discrete spaces~\cite{JEC}. A lattice of maximum entropy probability models is constructed based on marginal dependency constraints, and the unique information that a particular predictor has about the target is defined as the minimum increase in joint predictor-target mutual information when that particular predictor-target marginal dependency is constrained. 
Here, we apply the \Id approach to Gaussian systems, for which the marginally constrained maximum entropy models are Gaussian graphical models.  Closed form solutions for the \Id PID are derived for both univariate and multivariate Gaussian systems. Numerical and graphical illustrations are provided, together with practical and theoretical comparisons  of the \Id PID with the minimum mutual information partial information decomposition ($I_{\text{mmi}}$)~\cite{barrett}. In particular, it is proved that the \Imm method generally produces larger estimates of redundancy and synergy than does the \Id method. In discussion of the practical examples, the PIDs are complemented by the use of tests of deviance for the comparison of Gaussian graphical models.

\end{abstract}

\begin{flushleft}
\end{flushleft}

\section{Introduction}

The Partial Information Decomposition (PID)~\cite{WB} provides a theoretical framework to characterize and quantify the structure of multivariate information sharing. 
That is, given a \emph{target} variable $Y$, and a number of \emph{predictor} variables $X_i$ the PID attempts to describe the mutual information between the target and predictors $I(\{X_i\}; Y)$ in terms of that which is unique to each predictor, as well as that which is shared (redundant) or synergistic between subsets of predictors. 
However, while the PID framework provides a theoretical structure for this sharing, practical applications require measures to quantify the different terms.
Although a number of different candidate measures have been proposed, this remains an open area of research \cite{HSP, GK, BROJA, RI1, Chich, JEC, FinnLiz}. 

In~\cite{JEC} James et al.  recently proposed a measure based on dependency constraints, denoted $I_\text{dep}$, which quantifies the unique information conveyed by a single predictor.
In the case of two predictors, this is sufficient to obtain all four terms of the full PID; for higher order systems some terms remain indeterminate.
However, there are in any case a number of noted concerns with the PID approach for larger numbers of predictors, since a non-negative decomposition does not in general seem to be possible within the current framework \cite{BROJA1, RBOJ, RBOB, Rauh}. 

The $I_\text{dep}$ measure was derived and presented for discrete systems \cite{JEC}. 
However, there are many applications in which continuous variables might be subjected to the same analysis, and the PID approach has been considered for Gaussian systems \cite{barrett, RI1}.
\Id is derived from considering dependency constraints imposed within a lattice of maximum entropy probability models. 
Here, we apply the same logic to derive \Id in the case of continuous Gaussian variables.
In this case, the maximum entropy probability models are Gaussian graphical models~\cite{JW, DE, SLL}, also termed covariance selection models~\cite{AD}. We provide closed form expressions for the two predictor \Id PID, for both univariate and multivariate continuous Gaussian predictors and target.
Open-source matlab code implementing these measures is provided at
\url{https://github.com/robince/partial-info-decomp/blob/master/calc_pi_Idep_mvn.m}

First, we provide a brief review of the PID (Section 1.1) and the discrete \Id measure (Section 1.2).
In Section 2, we derive \Id for univariate Gaussian variables, and in Section 3 extend to multivariate Gaussian variables. 

\subsection{The Partial Information Decomposition}

The partial information decomposition was introduced in \cite{WB} as a method to decompose mutual information in a multivariate system in terms of redundancies and synergies within and between subsets of predictors. 
Formally, the PID is developed as the Mobi\"us inversion of a shared information measure over the lattice of antichains of predictor variables. We refer the reader to~\cite{WB}  for the full details.

In this manuscript, we focus on the case of two predictors, $X_0, X_1$, and a target $Y$. 
In this case, the mutual information $I(X_0,X_1; Y)$ is decomposed into four terms:
\begin{itemize}
    \item $\red$, the information about Y that is shared, common or redundant between $X_0$ and $X_1$, 
    \item $\unqA$, the information about Y that is available only from $X_0$,
    \item $\unqB$, the information about Y that is available only from $X_1$,
    \item $\syn$, the information about Y that is only available when $X_0$ and $X_1$ are observed together.
\end{itemize}

These terms satisfy the following intuitive relationships:
\begin{align}
I(X_0, X_1; Y) &= \red + \unqA + \unqB + \syn  \label{pid1}\\
I( X_0; Y) &= \red + \unqA \\
I( X_1;Y) &= \red + \unqB  \label{pid2}
\end{align}
Given the existence of these three constraints in terms of classical mutual information values, there is only one degree of freedom left to specify the bivariate PID.
With any of the four terms quantified, the remaining three can be easily calculated. 
The initial formulation of \cite{WB} was based on quantifying redundancy, and deriving the other quantities, but others have focussed on quantifying unique information or synergy directly. 

\subsubsection{The Partial Information Decomposition for Gaussian Variables}

The original definition of the PID and most of the subsequent work referenced above focussed on discrete variables. 
However, there are many applications where continuous-valued Gaussian variables are interesting subjects for information theoretic analysis. 
For example, simplified model systems~\cite{LPF, SS}  or empirical data analysis \cite{MWSLP, InceHBM}. 
In~\cite{barrett}, all discrete PID measures available at the time were considered and their principles applied to multivariate Gaussian systems, where one univariate component of the Gaussian is denoted the target.
It was shown~\cite{barrett} that for a univariate target, if $\red$, $\unqA$ and $\unqB$ depend only on the predictor-target marginal $(X_0,Y)$, $(X_1,Y)$ distributions, then there is a unique non-negative PID for which the redundancy is given by the minimum mutual information (MMI).
Several proposed discrete PID measures fall into this class \cite{WB, HSP, GK, BROJA, olbrich}, 
so for Gaussian systems these approaches are all equivalent and equal to the MMI PID. 
The full bivariate MMI PID is defined as follows:
\begin{align}
\text{red} & = \min\{ I(X_0; Y), I(X_1; Y)  \} \label{mmi1}\\
\text{unq0} &= \begin{cases}  0, & \text{if}\quad I(X_0;Y) < I(X_1;Y) \\
                   I(X_0;Y) -I(X_1;Y), & \text{otherwise} 
                   \end{cases} \\
\text{unq1} &= \begin{cases}  0, & \text{if}\quad I(X_1;Y) < I(X_0;Y) \\
                   I(X_1;Y) -I(X_0;Y), & \text{otherwise} 
                   \end{cases} \\
\text{syn} &= \begin{cases}  I(X_0, X_1;Y) - I(X_1;Y), & \text{if}\quad I(X_0;Y) < I(X_1;Y) \\
                   I(X_0, X_1;Y) -I(X_0;Y), & \text{otherwise} 
                   \end{cases}  \label{mmi4}
\end{align}
The MMI PID takes the redundancy component to be the minimum of the two mutual informations between the target and the predictors. Hence, one of the unique information components will always be zero. 

A more recent proposed measure, $I_\text{ccs}$, does not satisfy the Barrett conditions, as it depends also on the predictor-predictor marginal $(X_0,X_1)$ distribution and is therefore not equivalent to the MMI PID. 
While there is no closed form expression for $I_\text{ccs}$ for Gaussians, it has been implemented using Monte Carlo methods \cite{RI1}.
\Id \cite{JEC} is also not invariant to changes in the predictor-predictor marginal distribution, and therefore does not reduce to MMI in the Gaussian case either. 

\subsection{Unique Information via Dependency Constraints}
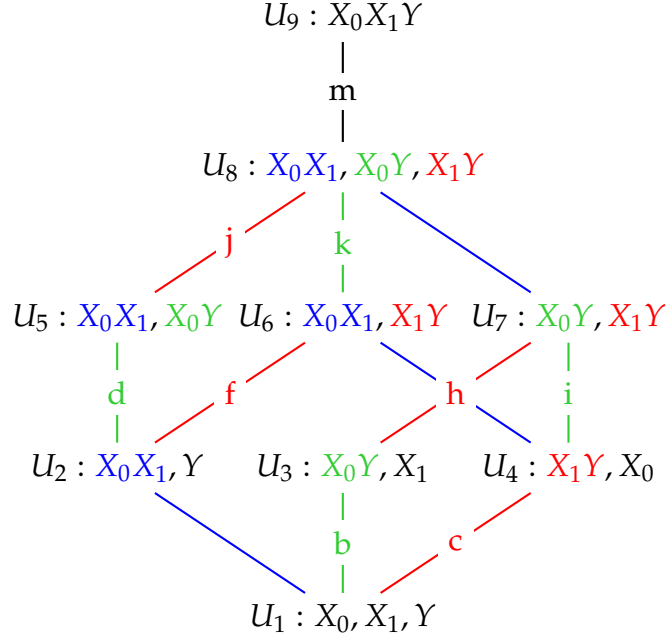
\begin{figure}[H]
      \begin{center}
            {\large

\begin{tikzpicture}
  \node (max) at (0,4) {$U_8: \textcolor{blue}{X_0X_1}, \textcolor{mygreen}{X_0Y}, \textcolor{red}{X_1Y}$};
  \node (a) at (-3,2) {$ U_5: \textcolor{blue}{X_0X_1}, \textcolor{mygreen}{X_0Y}  $};
  \node (b) at (0,2) {$ U_6: \textcolor{blue}{X_0X_1},  \textcolor{red}{X_1Y}$};
  \node (c) at (3,2) {$  U_7:  \textcolor{mygreen}{X_0Y}, \textcolor{red}{X_1Y}$};
  \node (d) at (-3,0) {$ U_2: \textcolor{blue}{X_0X_1}, Y$};
  \node (e) at (0,0) {$ U_3: \textcolor{mygreen}{X_0Y}, X_1$};
  \node (f) at (3,0) {$U_4:   \textcolor{red}{X_1Y}, X_0$};
  \node (min) at (0,-2) {$U_1: X_0, X_1, Y$};
    \node (three) at (0,6) {$U_9: X_0X_1Y$};
\draw[black, thick] (max) --(three)   node[midway, fill=white] {m} ;
\draw[blue, thick] (c) -- (max);
\draw[blue, thick] (f) -- (b);
\draw[blue, thick] (min) -- (d);

  \draw[mygreen, thick] (d) -- (a) node[midway, fill=white] {d};
   \draw[mygreen, thick] (min) -- (e) node[midway, fill=white] {b};
    \draw[mygreen, thick] (b) -- (max) node[midway, fill=white] {k};
  \draw [red, thick] (a) -- (max) node[midway, fill=white]{j};
  \draw[red, thick] (min) -- (f)  node[midway, fill=white] {c};
    \draw [red, thick] (d) -- (b) node[midway, fill=white]{f};
      \draw [red, thick] (e) -- (c) node[midway, fill=white]{h};
  \draw[mygreen, thick] (f) -- (c) node[midway, fill=white] {i};
\end{tikzpicture}
}
  \caption{A dependency lattice of models (based on~\cite{JEC}). Edges coloured green (b, d, i, k) correspond to adding the constraint $X_0Y$ to the model immediately below. Edges coloured red (c, f, h, j) correspond to adding the constraint $X_1Y$ to the model immediately below.  \label{fig1}}
      \end{center} 
      \end{figure}

In \cite{JEC} a method is proposed to quantify the unique information conveyed by a predictor variable. 
They start from a lattice of maximum entropy models subject to marginal constraints, where the lattice structure comes from the hierarchy of marginal constraints. 
This lattice is illustrated in Fig.~\ref{fig1}. 
For example, $U_1$ represents the maximum entropy distribution, having probability density function (p.d.f.) $g(x_0, x_1, y)$, under the constraints that the univariate marginals match exactly the univariate marginals of the original distribution, which has p.d.f. $f(x_0, x_1, y)$. That is: \linebreak $g(x_0) = f(x_0), \quad g(x_1) = f(x_1), \quad g(y) = f(y).$
$U_2$ represents the maximum entropy distribution subject to the constraints $g(x_0, x_1) = f(x_0, x_1), g(y) = f(y)$. $U_5$ represents the maximum entropy distribution subject to the constraints $ g(x_0, x_1) = f(x_0, x_1), g(x_0, y) = f(x_0, y)$, and so on.
The lattice structure arises from the higher order constraints enforcing corresponding lower order constraints, so that for example imposing a bivariate marginal constraint such as  $g(x_0, y) = f(x_0, y)$ means also that the lower order constraints $ g(x_0) = f(x_0)$ and $g(y) =f(y)$ also hold.
Note that in~\cite{JEC} there is  an additional model $U_9$ for the full distribution including third order interactions.  We focus here on Gaussian systems  which  are fully determined by their first and second order moments, and so do not feature any triple-wise interactions. Therefore, model $U_9$  doesn't appear in our lattice for Gaussian systems.

For the sake of brevity in the sequel, rather than speaking of imposing a constraint of the form  $g(x_0, y) = f(x_0, y)$, for example, we will speak of  'adding the constraint $X_0Y$'.

The colored edges correspond to adding a pairwise marginal constraint. 
Blue edges represent the constraint $X_0X_1$, i.e. preserving the pairwise dependency between $X_0$ and $X_1$. 
Green and red labelled edges correspond to the addition of the $X_0Y$ and the $X_1Y$ dependencies respectively.
For each model $U_1 \ldots U_8$ we calculate the mutual information between predictors and target under that model, $I_{U_i}( X_0, X_1;Y)$. 
The unique information in $X_0$ is then obtained as the minimum change in $I_{U_i}$ along all the green edges due to the addition of the  $X_0Y$ constraint to the model below.
Similarly, the unique information in $X_1$ can be obtained as the minimum change in $I_{U_i}$ along all the red edges due to the addition of the $X_1Y$ constraint to the model below. So, for example, the edge value $d$ is equal to $I_{U_5} - I_{U_2}$ and the edge value $f$ is equal to $I_{U_6} - I_{U_2}.$

If  the edge labels in Fig.~\ref{fig1} represent the change in mutual information along that edge, then the $I_\text{dep}$ PID is given by:
\begin{align}
    \text{unq0} &= \min\{b, d, i, k\}, \label{idep1}\\
    \text{red} &= I(X_0;Y) - \text{unq0}, \\
    \text{unq1} &= I(X_1;Y) - \text{red},  \\
    \text{syn} &= I(X_0,X_1;Y) - I(X_1;Y) - \text{unq0}  \label{idep4}
\end{align}
or:
\begin{align}
    \text{unq1} &= \min\{c, f, h, j\}, \\
    \text{red} &= I(X_0;Y) - \text{unq1}, \\
    \text{unq0} &= I(X_0;Y) - \text{red},\\
    \text{syn} &= I(X_0,X_1; Y) - I(X_0;Y) - \text{unq1}
\end{align}
It is shown in~\cite{JEC}  that this approach is consistent; the same PID results from either of the two forms above. 
They also show that the resulting PID satisfies the core axioms of symmetry, self-redundancy, monotonicity, local positivity and the identity axiom~\cite{JEC}.

\section{An \Id PID for univariate Gaussian predictors and target}

Since we will find in Section 2.2  that the required maximum entropy distributions that are described in Section 1.2 are Gaussian  graphical models we begin with a brief discussion of such models.
\subsection{Gaussian Graphical Models}

The independence graph for a probability distribution on three univariate random variables, $X_0, X_1, Y$ has three vertices and three possible edges, as described in Table~\ref{GM1}. Let $\mathbf{Z} = \begin{bmatrix} X_0 & X_1 &Y \end{bmatrix}^T$.

Graphical models represent the conditional independences present in a probability distribution, as described in Table 1.
Suppose that $\mathbf{Z}$ has a multivariate Gaussian distribution with  mean vector $\boldsymbol{\mu}_Z$, positive definite covariance matrix $\Sigma_{Z}$ and p.d.f. $f(x_0, x_1, y)$. There is no loss of generality in assuming that each component of $\mathbf{Z}$ has mean zero and variance equal to 1~\cite{barrett}. If we let the covariance (correlation) between $X_0$ and $X_1$ be $p$, between $X_0$ and $Y$ be $q$ and between $X_1$ and $Y$ be r, then the covariance (correlation) matrix for $\mathbf{Z}$ is
\begin{equation}
\Sigma{_Z} = \begin{bmatrix} 1 & p & q \\ p & 1 & r\\ q& r & 1  \label{cov1} \end{bmatrix} 
\end{equation}
and we require that $|p|, |q|, |r|$ are each less than 1, and to ensure positive definiteness we require also that $|\Sigma_Z| >0.$

Conditional independences are specified by setting certain off-diagonal entries to zero in the inverse covariance matrix, or concentration matrix, $K = \Sigma^{-1}$~\cite[p. 164]{JW}. Given our assumptions about the covariance matrix of $\boldmath{Z}$, this concentration matrix is
\begin{equation}
K =  \frac{1}{|\Sigma_Z|} \begin{bmatrix} 1 -r^2 & q r - p & p r - q \\ q r -p  & 1- q^2 & p q -r\\ p r -q& p q -r & 1- p^2 \end{bmatrix}, \label{K1}
\end{equation}
where $|\Sigma_Z| = 1 - p^2 - q^2 -r^2 + 2 p q r.$

We now illustrate using these Gaussian graphical models how  conditional independence constraints also impose constraints on marginal distributions of the type required, and we use the Gaussian graphical models $G_8$ and $G_6$ to do so. 
\begin{table}[H]
\centering
\caption{Graphical models and independences for the probability distribution of $\mathbf{Z}$. The vertices for random variables, $X_0, X_1, Y$ are denoted by $0, 1, 2$, respectively. Edges are denoted by pairs of vertices, such as $(1,2)$. In the column of independences, for example, $1 \indep 2 |0$ indicates that $X_1$ and $Y$ are conditionally independent given  $X_0$. (based on~\cite[p. 61]{JW})\label{GM1}}
\begin{tabular}{ccccl} \toprule 
Model &Independences & Edge Set  & Diagram & Description \\ \midrule
$G_1$ &  $1 \indep 2 \vert 0, \, 0 \indep 2| 1 $ & $\{ \}$  &  \begin{tikzpicture}[baseline= -5pt,
 roundnode/.style={circle, draw =black, fill =black, scale =0.3}]
\node[roundnode] at (0, 0) { }; \node[below] at (0, 0) {\small{0}};
\node[roundnode] at (.6, 0) {  };\node[below] at (.6, 0) {\small{1}};
\node[roundnode] at (1.2, 0) {  };\node[below] at (1.2, 0) {\small{2}};
\end{tikzpicture}

  & Mutual independence \\
 &  $ 1 \indep 2 | 0  $&   &   &\\\\
$G_2$ &  $2 \indep 0 \vert 1,\, 2 \indep 1 | 0$ & $\{(0,1)\}$ &  
\begin{tikzpicture}[baseline= -5pt,
 roundnode/.style={circle, draw =black, fill =black, scale =0.3}]
\node[roundnode] at (0, 0) { }; \node[below] at (0, 0) {\small{0}};
\node[roundnode] at (.6, 0) {  };\node[below] at (.6, 0) {\small{1}};
\node[roundnode] at (1.2, 0) {  };\node[below] at (1.2, 0) {\small{2}};
\draw[-] (0,0) -- (.6,0);
\end{tikzpicture}
 & Independent subsets \\

   $G_3$ & $1 \indep 0 \vert 2, \, 1 \indep 2 | 0$ & $\{(0,2)\}$ &\begin{tikzpicture}[baseline=-2pt,
 roundnode/.style={circle, draw =black, fill =black, scale =0.3}]
\node[roundnode] at (0, 0) { }; \node[below] at (0, 0) {\small{0}};
\node[roundnode] at (.6, 0) {  };\node[below] at (.6, 0) {\small{1}};
\node[roundnode] at (1.2, 0) {  };\node[below] at (1.2, 0) {\small{2}};
\draw[-] (0,0) .. controls (0.4, .5) and  (0.8, .5) .. (1.2,0);
\end{tikzpicture}
 & Independent subsets \\
 
$G_4$ &  $0 \indep 1 \vert 2, \, 0 \indep 2| 1 $& $\{(1,2)\}$ &\begin{tikzpicture}[baseline= -5pt,
 roundnode/.style={circle, draw =black, fill =black, scale =0.3}]
\node[roundnode] at (0, 0) { }; \node[below] at (0, 0) {\small{0}};
\node[roundnode] at (.6, 0) {  };\node[below] at (.6, 0) {\small{1}};
\node[roundnode] at (1.2, 0) {  };\node[below] at (1.2, 0) {\small{2}};
\draw[-] (.6, 0) -- (1.2,0);
\end{tikzpicture}

 & Independent subsets\\

  $G_5$ & $1 \indep 2 \vert 0$ & $\{(0,1), (0,2)\}$ &\begin{tikzpicture}[baseline=0pt,
 roundnode/.style={circle, draw =black, fill =black, scale =0.3}]
\node[roundnode] at (0, 0) { }; \node[below] at (0, 0) {\small{0}};
\node[roundnode] at (.6, .5) {  };\node[above] at (.6, .5) {\small{1}};
\node[roundnode] at (1.2, 0) {  };\node[below] at (1.2, 0) {\small{2}};
\draw[-] (0,0) -- (.6, .5);
\draw[-] (0, 0) -- (1.2, 0);

\end{tikzpicture} & One independence \\\\
 $G_6$ &  $0 \indep 2 \vert 1$ & $\{(0,1), (1,2)\}$ &\begin{tikzpicture}[baseline=0pt,
 roundnode/.style={circle, draw =black, fill =black, scale =0.3}]
\node[roundnode] at (0, 0) { }; \node[below] at (0, 0) {\small{0}};
\node[roundnode] at (.6, .5) {  };\node[above] at (.6, .5) {\small{1}};
\node[roundnode] at (1.2, 0) {  };\node[below] at (1.2, 0) {\small{2}};
\draw[-] (0,0) -- (.6,.5);
\draw[-] (1.2, 0) -- (.6, .5);
\end{tikzpicture}
 & One independence \\

    $G_7$ & $0 \indep 1 \vert 2 $& $\{(0, 2), (1,2)\}$ &\begin{tikzpicture}[baseline=0pt,
 roundnode/.style={circle, draw =black, fill =black, scale =0.3}]
\node[roundnode] at (0, 0) { }; \node[below] at (0, 0) {\small{0}};
\node[roundnode] at (.6, .5) {  };\node[above] at (.6, .5) {\small{1}};
\node[roundnode] at (1.2, 0) {  };\node[below] at (1.2, 0) {\small{2}};
\draw[-] (0,0) -- (1.2,0);
\draw[-] (1.2, 0) -- (.6, .5);
\end{tikzpicture}
 & One independence \\

 $G_8$&None &  $\{(0,1), (0,2), (1,2)\}$  & \begin{tikzpicture}[baseline=0pt, 
 roundnode/.style={circle, draw =black, fill =black, scale =0.3}] 
\node[roundnode] at (0, 0) { }; \node[below] at (0, 0) {\small{0}};
\node[roundnode] at (.6, .5) {  };\node[above] at (.6, .5) {\small{1}};
\node[roundnode] at (1.2, 0) {  };\node[below] at (1.2, 0) {\small{2}};
\draw[-] (0,0) -- (.6,.5);
\draw[-] (1.2, 0) -- (.6, .5);
\draw[-] (0,0) -- (1.2, 0);
\end{tikzpicture}
  &Complete interdependence \\

 \bottomrule
 \end{tabular}
 \end{table}

Since $\mathbf{Z}$ is multivariate Gaussian and has a zero mean vector, the distribution of $\mathbf{Z}$ is specified via its  covariance matrix $\Sigma_Z$. Hence, fitting any of the Gaussian graphical models $G_1 \ldots G_8$ involves estimating the relevant covariance matrix by taking the conditional independence constraints into account. Let $\hat{\Sigma}_i$ and $\hat{K_i}$ be the covariance and concentration matrices of the fitted model $G_i, \,\,(i =1 \ldots 8)$. 

We begin with the saturated model $G_8$ which has a fully connected graph and no constraints of conditional independence. Therefore, there is no need to set any entries of the concentration matrix $K$ to zero, and so $\hat{\Sigma}_8$ = $\Sigma_Z$. That is: model $G_8$ is equal to the given model for $\mathbf{Z}$.

Now consider model $G_6$.  In this model there is no edge between $X_0$ and $Y$ and so $X_0$ and $Y$ are conditionally independent given  $X_1$. This conditional independence is enforced by ensuring that the [1, 3] and [3, 1] entries in $\hat{K}_6$ are zero. The other elements in $\hat{K}_6$ remain to be determined. Therefore $\hat{K}_6$ has the form
 \begin{equation}
 \hat{K}_6 = \begin{bmatrix} \hat{k}_{00} & \hat{k}_{01} & 0 \\ \hat{k}_{01} & \hat{k}_{11} & \hat{k}_{12} \\ 0& \hat{k}_{12} & \hat{k}_{22} \end{bmatrix}.
 \label{conc1}
 \end{equation}
Given the form of $\hat{K}_6$ , $\hat{\Sigma}_6$ has  the form
 \begin{equation}
 \hat{\Sigma}_6 = \begin{bmatrix} 1 & p & \hat{\sigma}_{02} \\ p & 1 & r\\  \hat{\sigma}_{02} & r & 1  \end{bmatrix},
 \label{margcons}
 \end{equation} where $\hat{\sigma}_{02}$ is to be determined.  Notice that only the [1, 3] and [3, 1] entries in $\hat{\Sigma}_6$ have been changed from the given covariance matrix $\Sigma_Z$, since the [1,3] and [3, 1] entries of $\hat{K}_6$ have been set to zero.
 An exact solution is possible. The inverse of $\hat{\Sigma}_6$ is
 \begin{equation}
 \hat{K}_6 = \hat{\Sigma}_6^{-1} = \frac{1}{|\hat{\Sigma}_6|} \begin{bmatrix}  1 -r^2 & \hat{\sigma}_{02} r - p & p r - \hat{\sigma}_{02} \\  \hat{\sigma}_{02} r -p  & 1-  \hat{\sigma}_{02}^2 & p  \hat{\sigma}_{02} -r\\ p r - \hat{\sigma}_{02}& p  \hat{\sigma}_{02} -r & 1- p^2   \end{bmatrix}
 \label{solve1}
 \end{equation}
 Since the [1,3] entry in $\hat{K}_6$ must be zero, we obtain the solution that $\hat{\sigma}_{02} = p r,$ and so the estimated covariance matrix for model $G_6$ is
 \begin{equation}
  \hat{\Sigma}_6 = \begin{bmatrix} 1 & p & p r \\ p & 1 & r\\  p r& r & 1  \end{bmatrix}.
 \label{optcov1}
 \end{equation}
 The estimated covariance matrices for the other models can be obtained exactly using a similar argument.

 Model $G_6$ contains the marginal distributions of $X_0$, $X_1$, $Y$, $(X_0, X_1)$ and $(X_0, Y)$. It is important to note that these marginal distributions are exactly the same as in the given  multivariate Gaussian distribution for $\mathbf{Z}$, which has  covariance matrix $\Sigma_Z$. To see this we use a standard result on the marginal distribution of a sub-vector of a multivariate Gaussian distribution~\cite[p. 63]{MKB}. 
 
 The covariance matrix of the marginal distribution $(X_0, X_1)$ is equal to the upper-left 2 by 2 sub-matrix of $\hat{\Sigma}_6$, which is also equal to the same sub-matrix in $\Sigma_Z$ in~$\eqref{cov1}$. This means that this marginal distribution in model $G_6$ is equal to the corresponding marginal distribution in the distribution of $Z$. The covariance matrix of the marginal distribution $(X_0, Y)$ is equal to the lower-right 2 by 2 sub-matrix of $\hat{\Sigma}_6$, which is also equal to the same sub-matrix in $\Sigma_Z$ in~$\eqref{cov1}$, and so the $(X_0, Y)$ marginal distribution in model $G_6$ matches the corresponding marginal distribution in the distribution of $Z$. 
 Using similar arguments, such equality is also true for the other marginal distributions in model $G_6$.

 Looking at~$\eqref{K1}$, we see that setting to [1, 3] of $K$ entry to zero gives $q = p r$. Therefore, simply imposing this conditional independence constraint also gives the required estimated covariance matrix $\hat{\Sigma}_6$.

    It is generally true~\cite[p.176]{JW} that applying the conditional independence constraints is sufficient and it also leads to the marginal distributions in the fitted model being exactly the same as the corresponding marginal distributions in the given distribution of $\mathbf{Z}$. For example, in~$\eqref{margcons}$ we see that the only elements in $\hat{\Sigma}_6$ that are altered are the [1, 3]  and [3, 1] entries and these entries corresponds exactly to the zero [1, 3] and [3, 1] entries in $\hat{K}_6$. 
  That is: the location of zeroes in $\hat{K}_6$ determines which entries in $\hat{\Sigma}_6$ will be changed; the remaining entries of $\hat{\Sigma}_Z$ are unaltered and therefore this fixes the required marginal distributions. 
  Therefore, in Section 2.2, we will determine the required maximum entropy solutions by simply applying the necessary conditional independence constraints together with the other required constraints.

 We may express the combination of the constraints on marginal distributions and the constraints imposed by conditional independences as follows~\cite{AD}. For model $G_k$, the $(i, j)$th entry of $\hat{\Sigma}_Z$ is given by
 $$ \hat{\Sigma}_Z[i, j] = \Sigma_Z[i,j], \quad \text{for} \quad i=j \quad \text{and} \quad (i, j) \in E_k, $$
 where $E_k$ is the edge set for model $G_k$ (see Table 1). For model $G_k$, the conditional independences are imposed by setting the $(i, j)$th entry of $\hat{K}$ to zero whenever $(i, j) \not\in E_k.$
  
 Before moving on to derive the maximum entropy distributions, we consider the conditional independence constraints in model $G_3$. In model $G_3$ we see from Table~\ref{GM1} that this model has no edge between $X_0$ and $X_1$ and none between $X_1$ and $Y$. Hence, $X_0$ and $X_1$ are conditionally independent given $Y$ and also $X_1$ and $Y$ are conditionally independent given $X_0$. Hence, in $K$ in~$\eqref{K1}$ we set the [1, 2] and [2, 3] (and the [2, 1] \& [3, 2]) entries to zero to enforce these conditional independences. That is: $p= q r$ and $r = p q$. Taken together these equations give that $p=0$ and $r=0$, and so the  estimated covariance matrix for model $G_3$ is
  \begin{equation}
  \hat{\Sigma}_3 = \begin{bmatrix} 1 & 0 & q \\ 0 & 1 & 0\\  q& 0 & 1  \end{bmatrix}.
 \label{optcov3}
 \end{equation}
We also note that model $U_3$ in Fig.~\ref{fig1} also possesses the same conditional independences as $G_3$. This is true for all of the maximum entropy models $U_i$, and so when finding the nature of these models in the next section we apply in each case the conditional independence constraints satisfied by the graphical model $G_i$.

\subsection{Maximum Entropy distributions}

We are given the distribution of $Z$ which is multivariate Gaussian with zero mean vector and covariance matrix $\Sigma_Z$ in~$\eqref{cov1}$, and has p.d.f. $f(\mathbf{z}) \equiv f(x_0, x_1, y).$ For each of the models $U_1 \ldots U_8$, we will determine  the  p.d.f. of the maximum entropy solution $g(\mathbf{z}) \equiv g(x_0, x_1, y)$ subject to the constraints
\begin{equation}
\int_{\mathbb{R}^3} \mathbf{z} \,g(\mathbf{z}) \,\, d\mathbf{z} = \mathbf{0}, \quad \int_{\mathbb{R}^3} g(\mathbf{z}) \,\, d\mathbf{z}  = 1, \quad g(\mathbf{z}) > 0, \label{MEC1U}
\end{equation}
and the separate constraint for model $U_i$
\begin{equation}
\int_{\mathbb{R}^3}\mathbf{z z}^T g(\mathbf{z}) \,\, d \mathbf{z} = \hat{\Sigma}_i, \label{MEC2U}
\end{equation}
as well as the conditional independence constraints given in Table~\ref{CIcon1}.
\begin{table}[H]
\centering
\caption{Conditional independence constraints satisfied by the Gaussian graphical models $G_1 \ldots G_7$ that are applied when determining the maximum entropy models $U_1 \ldots U_7$.} \label{CIcon1}
\begin{tabular}{lll} \toprule
$U_1: p=q r, q = p r, r = p q $& & \\
$U_2: q = p r, r = p q $&  $U_3: p = q r, r = p q $& $U_4: p = q r, q = p r $ \\
$U_5:  r = p q $&  $U_6:  q = p r $& $U_7: p = q r $\\
\bottomrule
\end{tabular}
\end{table}
We begin with model $U_8$. As shown in the previous section, the estimated covariance matrix for model $U_8$, $\hat{\Sigma}_8$,  is equal to the covariance matrix of $\mathbf{Z}$, $\Sigma_Z$. By a well-known result~\cite{CT}, the solution is that $U_8$ is multivariate Gaussian with mean vector zero and covariance matrix, $\Sigma_Z$. That is: $U_8$ is equal to the given distribution of $\mathbf{Z}$.

\begin{table}[H]
\centering
\caption {Covariance matrices, with corresponding concentration matrices, for the Gaussian graphical models which were derived as maximum entropy probability models in Proposition 1. \label{MEcov1} }
\begin{tabular}{lcc} \toprule
Model & $\hat{\Sigma}_i$ & $\hat{K_i} $ \\ \midrule
$U_1: X_0, X_1, Y$ & $\begin{bmatrix} 1 & 0 & 0\\ 0 & 1 & 0\\ 0& 0& 1 \end{bmatrix} $ &  $\begin{bmatrix} 1 & 0 & 0\\ 0 & 1 & 0\\ 0& 0& 1  \end{bmatrix} $\\\\
$U_2: X_0X_1, Y$ & $\begin{bmatrix} 1 & p & 0\\ p & 1 & 0\\ 0& 0& 1 \end{bmatrix} $ &  $\frac{1}{1-p^2} \begin{bmatrix} 1 & -p & 0\\ -p & 1 & 0\\ 0& 0& 1 -p^2\end{bmatrix} $\\
$U_3: X_0Y, X_1$ & $\begin{bmatrix} 1 & 0 & q\\ 0 & 1 & 0\\ q& 0& 1 \end{bmatrix} $ &  $\frac{1}{1-q^2}\begin{bmatrix} 1 & 0 & -q\\ 0 & 1-q^2 & 0\\ -q& 0& 1 \end{bmatrix} $\\
$U_4: X_1Y, X_0$ & $\begin{bmatrix} 1 & 0 & 0\\ 0 & 1 & r\\ 0& r& 1 \end{bmatrix} $ &  $ \frac{1}{1-r^2}\begin{bmatrix} 1-r^2 & 0 & 0\\ 0 & 1 & -r\\ 0& -r& 1 \end{bmatrix} $\\\\
$U_5: X_0X_1, X_0Y$ &   $\begin{bmatrix} 1 & p & q\\ p & 1 & pq\\ q& pq& 1 \end{bmatrix} $ &  $ \frac{1}{(1-p^2)(1-q^2)} \begin{bmatrix} 1-p^2q^2 & (q ^2 -1)p &  (p ^2 -1)q \\ (q^2 -1)p& 1-q^2 & 0 \\ (p^2 -1)q & 0 & 1-p^2 \end{bmatrix} $\\\\
$U_6: X_0X_1, X_1Y$ &  $\begin{bmatrix} 1 & p & p r\\ p & 1 & r\\ p r& r& 1 \end{bmatrix} $ &  $ \frac{1}{(1-p^2)(1-r^2)} \begin{bmatrix} 1-r^2 & (r^2 -1)p &  0 \\ (r^2 -1)p & 1-p^2r^2 & (p^2-1)r \\ 0 & (p^2 -1)r& 1-p^2 \end{bmatrix} $\\\\
$U_7:X_0Y, X_1Y$ & $\begin{bmatrix} 1 & q r & q\\  q r  & 1 & r\\ q& r& 1 \end{bmatrix} $ &  $ \frac{1}{(1-q^2)(1-r^2)}\begin{bmatrix} 1-r^2 & 0 &  (r^2-1)q\\ 0 & 1-q^2 & (q^2-1)r\\ (r^2 -1)q& (q^2 -1)r& 1-q^2r^2 \end{bmatrix} $\\\\
$U_8:X_0X_1, X_0Y, X_1Y$ & $\begin{bmatrix} 1 & p & q\\ p & 1 & r\\ q& r& 1 \end{bmatrix} $ &  $ \frac{1}{|\Sigma_Z|} \begin{bmatrix} 1-r^2 & q r -p &  p r -q\\ q r -p & 1-q^2 & p q -r\\ p r -q& p q -r& 1-p^2 \end{bmatrix} $\\
\bottomrule
\end{tabular}
\end{table}

For model $U_5$, the conditional independence constraint is $r = p q$ and so
 \begin{equation}
  \hat{\Sigma}_5 = \begin{bmatrix} 1 & p & q \\ p & 1 & p q\\  q& p q & 1  \end{bmatrix}.
 \label{optcov5}
 \end{equation}
Hence, using a similar argument to that for $U_8$, the maximum entropy solution for model $U_5$ is multivariate Gaussian with zero mean vector and covariance matrix $\hat{\Sigma}_5$, and so is equal to the model $G_5$.

In model $U_3$, the conditional independence constraints are $p= q r, r =p q$ and so $p=0$ and $r=0$. Therefore, 
\begin{equation}
  \hat{\Sigma}_3 = \begin{bmatrix} 1 & 0 & q \\ 0 & 1 &  0\\  q& 0 & 1  \end{bmatrix}
 \label{optcov32}
 \end{equation}
and the maximum entropy solution for $U_3$ is multivariate Gaussian with zero mean vector and covariance matrix $\hat{\Sigma}_3$, and so is equal to $G_3$.  The derivations for the other maximum entropy models are similar, and we state the results in Proposition 1.
\begin{proposition}
The distributions of maximum entropy, $U_1 \ldots U_8$, subject to the constraints~$\eqref{MEC1U}$-$\eqref{MEC2U}$ and the conditional independence constraints in Table~\ref{CIcon1},
are trivariate Gaussian  graphical models $G_1 \ldots G_8$ having mean vector $\mathbf{0}$ and with the covariance matrices $\hat{\Sigma}_i, \,\,(i=1, \ldots 8),$  given above in Table~\ref{MEcov1}
\end{proposition}
The estimated covariance  matrices in Table~\ref{MEcov1} were inverted to give the corresponding concentration matrices, which are also given in Table~\ref{MEcov1}. They indicate by the location of the zeroes that the conditional independences have been appropriately applied in the derivation of the results in Proposition 1.

It is important to check that the relevant bivariate and univariate marginal distributions are the same in all of the models in which a particular constraint has been added For example, the $X_0X_1$ constraint is present in models $U_2, U_5, U_6, U_8.$ The marginal bivariate $X_0X_1$ distribution has zero mean vector and so is determined by the upper-left  2  by 2 sub-matrix of the estimated covariance matrices, $\hat{\Sigma}_i$~\cite[p. 63]{MKB}. Inspection of Table~\ref{MEcov1} shows that this sub-matrix is equal to $  \begin{bmatrix} 1 & p \\ p & 1 \end{bmatrix}$ in all four models. Thus the bivariate distribution of $(X_0, X_1)$ is the same in all four models in which this dependency constraint is fitted. It is also the same as in the original distribution, which has covariance matrix $\Sigma_Z$ in~$\eqref{cov1}$. Further examination of Table~\ref{MEcov1} shows equivalent results for the $(X_0,Y)$ and $(X_1,Y)$ bivariate marginal distributions. The univariate term $Y$ is present in all eight models. The univariate distribution of $Y$ has mean zero and so is determined by the [3,3] element of the estimated covariance matrices $\hat{\Sigma}_i$~\cite[p. 63]{MKB}. Looking at the $\hat{\Sigma}_i$ column, we see that the variance of $Y$ is equal to 1 in all eight models, and so the marginal distribution of $Y$ is the same in all eight models. In particular, this  is true in the original distribution, which has covariance matrix $\Sigma_Z$ in~$\eqref{cov1}$.

\subsection{Mutual Information}
Some required results involving  mutual information will now be stated. They will be used to find expressions for the total mutual information of each model and also in constructing the \Id and \Imm PIDs.
\begin{align}
I(X_0, X_1; Y) &=  \frac{1}{2} \log \left(\frac{1- p^2}{1 - p^2 -q^2 -r^2 + 2 p q r} \right), \label{detsig} \\
I(X_0; Y) &= \frac{1}{2} \log \left( \frac{1}{1-q^2}  \right), \label{bivMI0}\\
I(X_1; Y) & = \frac{1}{2} \log \left( \frac{1}{1-r^2}  \right), \label{bivMI1} 
\end{align}

\begin{align}
I(X_0;Y|X_1) &= \frac{1}{2} \log \left(  \frac{(1-p^2)(1-r^2)}{1 - p^2 -q^2 -r^2 + 2 p q r}  \right), \label{condMI1}\\
I(X_1;Y|X_0) &= \frac{1}{2} \log \left(  \frac{(1-p^2)(1-q^2)}{1 - p^2 -q^2 -r^2 + 2 p q r}  \right). \label{condMI2}
\end{align}
Application of~$\eqref{detsig}$ with the covariance matrices given in Table~\ref{MEcov1} gives the following expressions for the total mutual information $I(X_0, X_1; Y)$ for the maximum entropy models derived in Proposition 1.

\setlength{\tabcolsep}{4em}
\begin{table}[H]
\centering
\caption{Expressions for the predictors-target mutual information for the eight models in the dependency lattice of Fig.~\ref{fig0}, as described in Table~\ref{MEcov1}. \label{MItots1}}
\begin{tabular}{ll} \toprule
$U_8:  \frac{1}{2} \log \left(\frac{1- p^2}{1 - p^2 -q^2 -r^2 + 2 p q r} \right) $& $U_4: I(X_1; Y)$\\\\
$U_7:  \frac{1}{2} \log \left(\frac{1- q^2r^2}{(1  -q^2)(1-r^2)} \right) $&$U_3: I(X_0; Y) $ \\\\
$U_6:  I(X_1;Y) = \frac{1}{2} \log \left(\frac{1}{1  -r^2} \right) $& $U_2: 0$\\\\
$U_5:  I(X_0; Y) =  \frac{1}{2} \log \left(\frac{1}{1 - q^2} \right) $&$U_1: 0 $\\

\bottomrule
\end{tabular}
\end{table}

\subsection{The \Id PID for univariate Gaussian predictors and target }

The \Id PID for Gaussian predictors and target will now be constructed; for details, see~$\eqref{idep1}$-$\eqref{idep4}$ in Section 1.2.
\begin{figure}[H]
      \begin{center}
            {\large
\begin{tikzpicture}
  \node (max) at (0,4) {$U_8: \textcolor{blue}{X_0X_1}, \textcolor{mygreen}{X_0Y}, \textcolor{red}{X_1Y}$};
  \node (a) at (-3,2) {$ U_5: \textcolor{blue}{X_0X_1}, \textcolor{mygreen}{X_0Y}  $};
  \node (b) at (0,2) {$ U_6: \textcolor{blue}{X_0X_1},  \textcolor{red}{X_1Y}$};
  \node (c) at (3,2) {$  U_7:  \textcolor{mygreen}{X_0Y}, \textcolor{red}{X_1Y}$};
  \node (d) at (-3,0) {$ U_2: \textcolor{blue}{X_0X_1}, Y$};
  \node (e) at (0,0) {$ U_3: \textcolor{mygreen}{X_0Y}, X_1$};
  \node (f) at (3,0) {$U_4:   \textcolor{red}{X_1Y}, X_0$};
  \node (min) at (0,-2) {$U_1: X_0, X_1, Y$};
  \draw[mygreen, thick] (d) -- (a) node[midway, fill=white] {d};
   \draw[mygreen, thick] (min) -- (e) node[midway, fill=white] {b};
    \draw[mygreen, thick] (b) -- (max) node[midway, fill=white] {k};
  \draw [red, thick] (a) -- (max) node[midway, fill=white]{j};
  \draw[red, thick] (min) -- (f)  node[midway, fill=white] {c};
    \draw [red, thick] (d) -- (b) node[midway, fill=white]{f};
      \draw [red, thick] (e) -- (c) node[midway, fill=white]{h};
  \draw[mygreen, thick] (f) -- (c) node[midway, fill=white] {i};
\end{tikzpicture}
}
  \caption{A dependency lattice of models (based on~\cite{JEC}.). Edges coloured green (b, d, i, k) correspond to adding the term $X_0Y$ to the model immediately below. Edges coloured red (c, f, h, j) correspond to adding the term $X_1Y$ to the model immediately below. The two relevant sub-lattices are shown here. \label{fig0}}
      \end{center} 
      \end{figure}
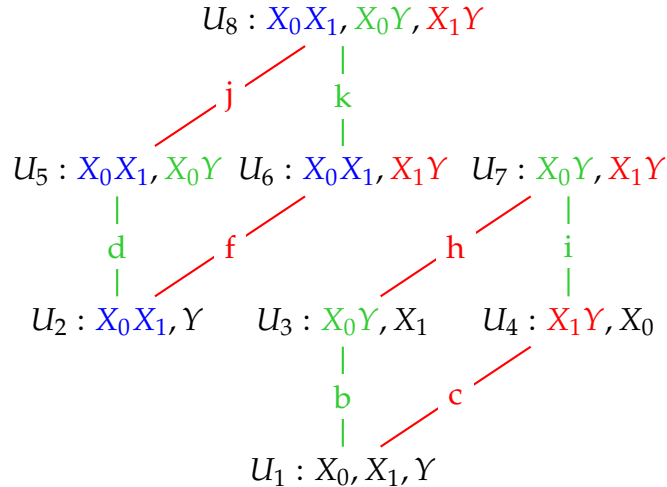
\noindent Using  the results in Table~\ref{MItots1} together with the dependency lattice in Fig. 2, we may write down expressions for all the required edge values, and they are given in Table~\ref{edges1}.

 \setlength{\tabcolsep}{2em}
\begin{table}[H]
\centering
\caption{Expression for the edge values in the dependency lattice in Fig. 2 that are used to determine the unique informations. \label{edges1}}
\begin{tabular}{ll} \toprule
$ b = I(X_0; Y) = \frac{1}{2} \log \left( \frac{1}{1- q^2}  \right)$ &  $ c = I(X_1; Y) =  \frac{1}{2} \log \left( \frac{1}{1- r^2}  \right)$\\\\
$d = I(X_0;Y) $&$f = I(X_1; Y) $ \\\\
$i = \frac{1}{2} \log \left( \frac{1 - q^2 r^2   } { (1-q^2)(1-r^2)  }  \right) - I(X_1; Y)   $& $ h = \frac{1}{2} \log \left( \frac{1 - q^2 r^2   } { (1-q^2)(1-r^2)  }  \right) - I(X_0; Y)    $\\\\
$ k = \frac{1}{2} \log \left( \frac{1 - p^2   } { 1- p^2 -q^2 -r^2 +2 p q r  }  \right) - I(X_1; Y)    $&$ j = \frac{1}{2} \log \left( \frac{1 - p^2   } { 1- p^2 -q^2 -r^2 +2 p q r  }  \right) - I(X_0; Y)     $\\\\

\bottomrule
\end{tabular}
\end{table}
\noindent By making use of the edge values given in Table~\ref{edges1} together with $\eqref{idep1}$-$\eqref{idep4}$ from Section 1.2 the \Id PID can be constructed.
We now state some results for the \Id PID for univariate Gaussian predictors, $X_0, X_1$, and target, $Y$, with proofs given in  Appendix A.
\begin{proposition} For two univariate Gaussian predictors, $X_0, X_1$, and one univariate Gaussian  target, $Y$, the \Id PID, defined in Table~\ref{edges1}, and $\eqref{idep1}$-$\eqref{idep4}$ in Section 1.2,   has the following properties.
\begin{itemize}
\item[(a)] The \Id PID possesses  consistency as well as the core axioms of non-negativity,  self-redundancy, monotonicity, symmetry and identity.
\item[(b)] When unq0 is equal to $b$ or $d$, the the redundancy component is zero.
\item[(c)] When unq0 is equal to $i$, the redundancy and both unique informations are constant with respect to the correlation between the two predictors.
\item[(d)] When the correlations between each predictor and the target are both non-zero, then unq0 is equal to either $i$ or to $k$.

\item[(e)] When unq0 is equal to $k$, the synergy component is zero.
\item[(f)] The redundancy component in the \Imm PID is greater than or equal to the redundancy component in the \Id PID with equality if, and only if, at least one of the following conditions holds: (i) either predictor and the target are independent; (ii) either predictor is conditionally independent of the target given the other predictor.
\item[(g)] The synergy component in the \Imm PID is greater than or equal to the synergy  component in the \Id PID with equality if, and only if,  at least one of the following conditions holds: (i) either predictor and the  target are independent; (ii) either predictor is conditionally independent of the target given the other predictor.
\item[(h)] The \Id and \Imm PIDs are identical when either $X_0$ and $Y$ are conditionally independent given $X_1$ or  $X_1$ and $Y$ are conditionally independent given $X_0$, and in particular they are identical for models $U_1 \ldots U_6$.  In model $U_7$ the synergy component of \Id is zero.

\end{itemize}

\end{proposition}

The \Imm PID is defined in $\eqref{mmi1}$-$\eqref{mmi4}$ in Section 1.1. We now consider examples of the \Id PID as well as comparisons between the \Imm  and \Id PIDs in the following subsections.

\subsection{Some examples}
\begin{example}
We consider the \Id PID when $q =\text{corr}(X_0, Y) =0, r \neq0, p \neq 0.$
\end{example}
When $q=0$, we see from Table 4 that $b=d=i=0$ and $k>0$, so unq0 =0, and since $I(X_0;Y) =0$ the redundancy component is also zero. The unique information, unq1, and the synergy component, syn, are  equal to 
$$I(X_1;Y) = \frac{1}{2} \log \frac{1}{1-r^2}, \quad  I(X_0;Y|X_1)= \frac{1}{2} \log \left(\frac{(1-p^2)(1-r^2)}{1 -p^2 -r^2} \right), $$
respectively. The \Imm PID is exactly the same as the \Id PID.
\begin{example}
We consider the \Id PID when $r =\text{corr}(X_1, Y) =0, q \neq0, p \neq 0.$
\end{example}
When $r=0$, we see from Table~\ref{edges1} that
$$ b = d = i = \frac{1}{2} \log \frac{1}{1-q^2} $$ and also that $k > \{b, d, i\}$ because
$$ (1-p^2)(1-q^2) > 1 -p^2 -q^2,$$ since $p \neq 0, q \neq 0.$ It follows that $\text{unq0} = \frac{1}{2} \log \frac{1}{1 - q^2}$, and that 
the synergy component is equal to $$ \frac{1}{2} \log \left(\frac{(1-p^2)(1-q^2)}{1 -p^2 -q^2} \right). $$ Since $I(X_1;Y) =0$, from~$\eqref{bivMI1}$,  the redundancy component is zero, as is unq1. The \Imm PID is exactly the same as the \Id PID.
\begin{example}
We consider the \Id PID when $p =\text{corr}(X_0, X_1) =0, q \neq0, r \neq 0.$
\end{example}
Under the stated conditions, it is easy to show that $ b < i$ and $i< k$ and so the minimum edge value is attained at $i$. Using the results in Table~\ref{edges1} and~$\eqref{bivMI1}$-$\eqref{condMI2}$, we may write down the \Id PID as follows.
\begin{align*}
\text{unq0} &=  \frac{1}{2} \log \left( \frac{1- q^2 r^2}{1 - q^2}  \right)\\
\text{unq1} &= \frac{1}{2} \log \left( \frac{1- q^2 r^2}{1 - r^2}  \right)\\
\text{red} &=  \frac{1}{2} \log \left( \frac{1}{1 - q^2}  \right) -  \frac{1}{2} \log \left( \frac{1- q^2 r^2}{1 - q^2}  \right) = \frac{1}{2} \log \left( \frac{1}{1- q^2 r^2} \right) \\
\text{syn} &= I(X_0; Y|X_1) - \text{unq0} =  \frac{1}{2} \log \left(  \frac{(1-q^2)(1-r^2)}{(1-q^2 -r^2)(1- q^2r^2)}       \right)
\end{align*}
For this situation, the \Imm PID takes two different forms, depending on whether or not $|q| < |r|$. Neither form is the same as the \Id  PID.
\begin{example} Compare the \Imm and \Id  PIDs when $p =-0.2$, $q=0.7$ and $r =-0.7$.
\end{example}
The PIDs are given in the following table.
\begin{center}
\begin{tabular}{ccccc} \toprule
PID & unq0 & unq1 & red & syn \\ \midrule
\Id & \W0.2877 & 0.2877 & 0.1981 & 0.4504 \\
\Imm & 0 & 0  & 0.4587 & 0.7380 \\

\bottomrule
\end{tabular}
\end{center}

There is a stark contrast between the two PIDs in this system. Since $|q|=|r|$, the \Imm PID has two zero unique informations, whereas \Id has equal values for the uniques but they are quite large. The \Imm PID gives much larger values for the redundancy and synergy components than does the \Id PID. In order to explore the differences between these PIDs, 50 random samples were generated from a multivariate normal distribution having correlations $ p=-0.2, q=0.7, r=-0.7.$ The sample estimates of $p, q, r$ were
$\hat{p} = -0.1125, \hat{q} = 0.6492, \hat{r} =-0.6915$ and the sample PIDs are 
\begin{center}
\begin{tabular}{ccccc} \toprule
PID & unq0 & unq1 & red & syn \\ \midrule
\Id & \W0.2324 & 0.3068 & 0.1623 & 0.4921 \\
\Imm & 0 & 0.0744  & 0.3948 & 0.7245 \\

\bottomrule
\end{tabular}
\end{center}
We now apply tests of deviance  in order to test model $U_i$ within the saturated model $U_8$. The null hypothesis being tested is that model $U_i$ is true (see Appendix E). 
The results of applying  tests of deviance~\cite[p. 185]{JW}, in which each of models $U_1 \ldots U_7$ is tested against the  saturated model $U_8$, produced  approximate $p$ values that were  close to zero  ($ p < 10^{-11}$) for all but model $U_7,$ which had a $p$ value of $3 \times 10^{-5}$. This suggests that none of the models $U_1 \ldots U_7$ provides an adequate fit to the data and so model $U_8$ provides the best description. The results of testing $U_6$ and $U_7$ within model $U_8$ gave strong evidence to suggest that the interaction terms $X_0Y$ and  $X_0Y$ are required to describe the data, and that each term makes a significant contribution in addition to the presence of the other term. Therefore, one would expect to find fairly sizeable unique components in a PID, and so the \Id PID seems to provide a more sensible answer in this example. One would also expect synergy to be present, and both PIDs have a large, positive synergy component.

\begin{example} Prediction of grip strength
\end{example}
Some data concerning the prediction of grip strength from physical measurements was collected from 84 male students at Glasgow University.  Let $Y$ be the grip strength, $X_0$ be the bicep circumference and $X_1$ the forearm circumference.  The following correlations between each pair of variables were calculated: $\text{corr}(X_1, Y) =0.7168, \text{corr}(X_0, Y) = 0.6383, \text{corr}(X_0, X_1) = 0.8484,$ and PIDs applied with the following results.
\begin{center}
\begin{tabular}{ccccc} \toprule
PID & unq0 & unq1 & red & syn \\ \midrule
\Id & \W0.0048 & 0.1476 & 0.3726 & 0 \\
\Imm & 0 & 0.1427 & 0.3775 & 0.0048 \\
\bottomrule
\end{tabular}
\end{center}
The \Id and \Imm PIDs are very similar, and the curious fact that unq0 in \Id is equal to the synergy in \Imm is no accident,  It is easy to show this connection theoretically by examining the results in~~$\eqref{condMI1}$-$\eqref{condMI2}$ and Table~\ref{edges1}; that is, the sum of unq0 and syn in  the \Id PID or the sum of unq1 and syn in the \Id PID is equal to the synergy value in the \Imm PID. This happens because the \Imm PID must have a zero unique component.

These PIDs indicate that there is almost no synergy among the three variables, which makes sense because the value of $I(X_0;Y|X_1)$ is close to zero, and this suggests that $X_0$ and $Y$ are conditionally independent given  $X_1$. On the other hand, $I(X_1;Y|X_0)$ is 0.1427 which suggests that $X_1$ and $Y$ are not conditionally independent given $X_0$, and so both terms $X_0X_1$ and $X_0Y$ are of relevance in explaining the data, which is the case in model $U_6.$ This model has  $I(X_0;Y|X_1)=0$  and therefore no synergy and also a zero unique value in relation to $X_0$. The results of applying tests of deviance~\cite[p. 185]{JW}, in which each of models $U_1 \ldots U_7$ is tested within the  saturated model $U_8$, show that the approximate $p$ values  are close to zero ($ p < 10^{-14}$) for all models except $U_5$ and $U_6$. The $p$ value for the model $U_5$ is  $3 \times 10^{-5}$, while the $p$ value for the test of $U_6$ against $U_8$ is  approximately 0.45. Thus there is strong evidence to reject all the models except model $U_6$ and  this suggests that model $U_6$ provides a good fit to data, and this alternative viewpoint provides support for the form of both PIDs.

\subsection{Graphical illustrations}

\begin{figure}[H]
\centering
 \setlength{\tabcolsep}{0.5em}
\begin{tabular}{cc}
\begin{subfigure}{0.4\textwidth}\centering\includegraphics[scale=0.5]{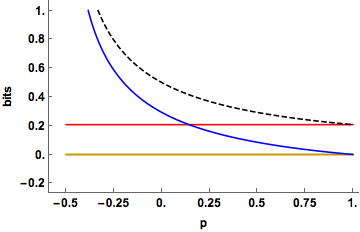}
\caption{$ \text{\Imm}, q = 0.5, r =0.5$} \label{MMIU1} \end{subfigure} &\begin{subfigure}{0.4\textwidth}\centering\includegraphics[scale=0.5]{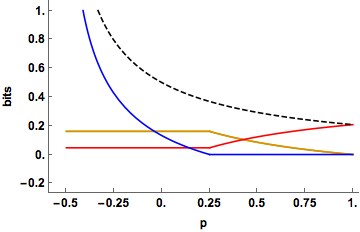}
\caption{$ \text{\Id}, q = 0.5, r=0.5$}  \label{IdepU1}\end{subfigure} \\
\end{tabular}
\begin{center}
\includegraphics[scale=0.5]{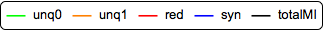} 
\end{center}
\begin{tabular}{cc}
\begin{subfigure}{0.4\textwidth}\centering\includegraphics[scale=0.5]{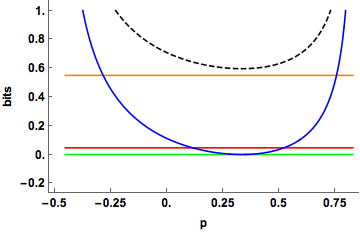}
\caption{$ \text{\Imm}, q =0.25, r = 0.75$}  \label{MMIU2}\end{subfigure}&
\begin{subfigure}{0.4
\textwidth}\centering\includegraphics[scale=0.5]{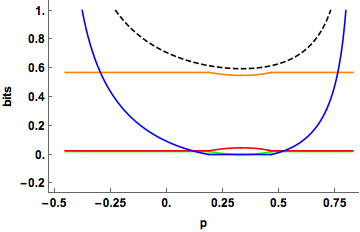}
\caption{$ \text{\Id}, q = 0.25, r = 0.75$}  \label{IdepU3}\end{subfigure}
\end{tabular}
 \caption{The \Imm \& \Id PID components are plotted for a range of values of the correlation ($p$) between the two predictors. Two combinations of the correlations ($q, r$) between each predictor and the target are displayed. The total mutual information $I(X_0, X_1; Y)$ is also shown as a dashed black curve.   } \label{fig3}
\end{figure}

We present some graphical illustrations of the \Id PID and compare it to the  \Imm PID; see sections 1.1.1 \& 1.2 for definitions of these PIDs.

Since $q=r$, both the \Imm unique informations are zero in Fig.~\ref{MMIU1}. The redundancy component is constant, while the synergy component decreases towards zero. In Fig.~\ref{IdepU1}, we observe change-point behaviour of \Id when $p=0.25$. For $p<0.25$ the unique components of \Id are equal, constant  and positive. The redundancy component is also constant and positive with a lower value than the corresponding component in the \Imm PID. The synergy component decreases towards zero and reaches this value when $p=0.25$. The \Id synergy is lower that the corresponding \Imm synergy for all values of $p$.

At $p=0.25$, the synergy "switches off" in the \Id  PID, and stays "off" for larger values of $p$,  and then the unique and redundancy components are free to change. In the range $0.25 < p < 1,$ the redundancy increases and takes up all the mutual information when $p=1$, while the unique informations decrease towards zero. The \Id and \Imm profiles show different features in this case. The "regime switching" in the \Id PID is interesting. As mentioned in Proposition 2, the minimum edge value occurs with unq0 =$i$ or $k$. When unq0=$k$ the synergy must be equal to zero, whereas when unq0=$i$ the synergy is positive and the values of the unique informations and the redundancy are constant. Regions of zero synergy in the \Id PID are explored in Fig.~\ref{ZS}.

\begin{figure}[H]
\centering

\begin{tabular}{cc}
\begin{subfigure}{0.4\textwidth}\centering\includegraphics[scale=0.5]{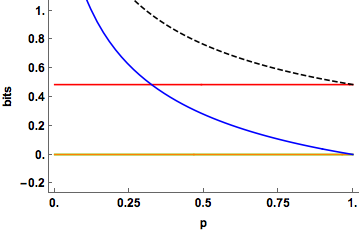}
\caption{$ \text{\Imm}, q = 0.7, r =0.7$} \label{MMIU4} \end{subfigure} &\begin{subfigure}{0.4\textwidth}\centering\includegraphics[scale=0.5]{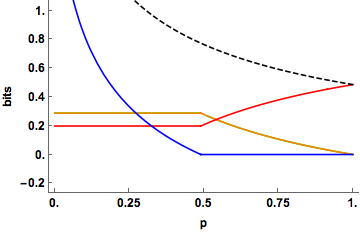}
\caption{$ \text{\Id}, q = 0.7, r= 0.7$} \label{IdepU4}\end{subfigure} \\
\end{tabular}
\begin{center}
\includegraphics[scale=0.5]{IdepLegend.png} 
\end{center}
\begin{tabular}{cc}
\begin{subfigure}{0.4\textwidth}\centering\includegraphics[scale=0.5]{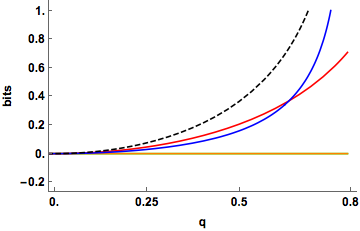}
\caption{$ \text{\Imm}, p=0.25$}  \label{MMIU3}\end{subfigure}&
\begin{subfigure}{0.4\textwidth}\centering\includegraphics[scale=0.5]{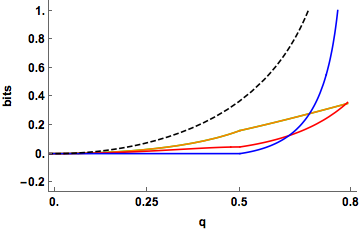}
\caption{$ \text{\Id}, p=0.25$} \label{IdepU2} \end{subfigure}
\end{tabular}
 \caption{In (a) \& (b), the \Imm \& \Id PIDs are plotted for a range of values of the correlation ($p$) between the two predictors. One combination of the correlations ($q, r$) between each predictor and the target are displayed. In (c) \& (d), the \Imm \& \Id PID are plotted for a range of allowable values of $q$, where $q$ is equal to $r$, for  $p=0.25$. The total mutual information $I(X_0, X_1; Y)$ is also shown as a dashed black curve.   } \label{fig4}
\end{figure}

In Figs.~\ref{MMIU4}-\ref{IdepU4}, there are clear differences in the PID profiles between the two methods. The \Id synergy component switches off at $p=0.5$ and is zero thereafter. For $p<0.5$, both the \Id uniques are much larger than those of \Imm, which are zero, and \Imm has a larger redundancy component. For $p>0.5$, the redundancy component in \Id increases to take up all of the mutual information, while the unique information components decrease towards zero. In contrast to this, in the \Imm PID the redundancy and unique components remain at their constant values while the synergy continues to decrease towards zero.

The PIDs are plotted for increasing values of $q=r$ in Figs.~\ref{MMIU3}-\ref{IdepU2} when $p=0.25$. The \Imm and \Id profiles are quite different. As $q$ increases, the \Imm uniques remain at zero, while the \Id uniques rise gradually. Both the \Imm redundancy and synergy profiles rise more quickly than their \Id counterparts, probably because both their uniques are zero. In the \Id PID, the synergy switches on at $p=0.5$ and it is noticeable than all the \Id components can change simultaneously as $q$ increases.

\begin{figure}[H]
\centering
\begin{tabular}{cc}
\begin{subfigure}{0.4\textwidth}\includegraphics[scale=0.5]{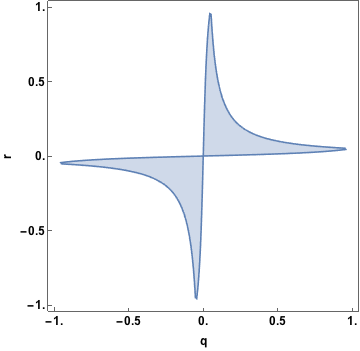}
\caption{$p = 0.05$} \end{subfigure} &\begin{subfigure}{0.4\textwidth}\includegraphics[scale=0.5]{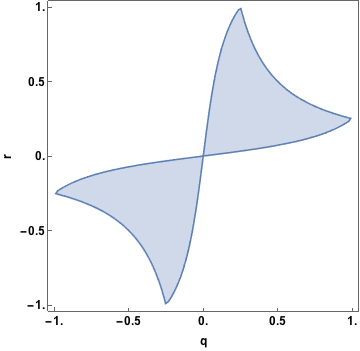}
\caption{$ p =0.25$} \end{subfigure} \\\\
\begin{subfigure}{0.4\textwidth}\includegraphics[scale=0.5]{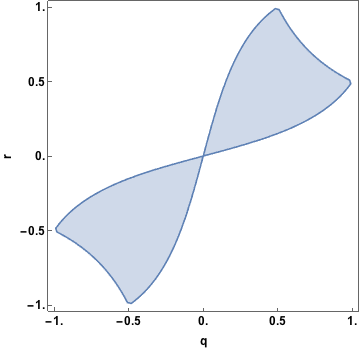}
\caption{$ p=0.5$}  \end{subfigure}&
\begin{subfigure}{0.4\textwidth}\includegraphics[scale=0.5]{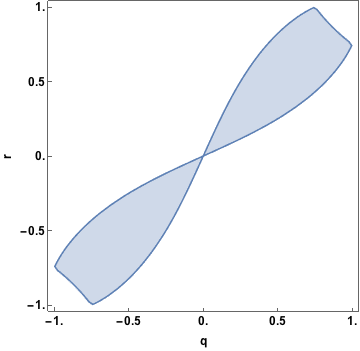}
\caption{$ p= 0.75$} \end{subfigure}
\end{tabular}
 \caption{Regions in $(q, r)$ space in which the synergy component in the \Id PID is equal to zero, plotted for four different values of $p$. Also, the determinant of $\Sigma_Z$ is positive.  } \label{ZS}
\end{figure}

One of the characteristics noticed in Figs.~\ref{fig3}-\ref{fig4}  is the 'switching behaviour' of the \Id PID in that there are kinks in the plots of the PIDs against the correlation between the predictors, $p$: the synergy component abruptly becomes equal to zero at certain values of $p$, and there are other values of $p$ at which the synergy moves from being zero to being positive.

 In Proposition 2, it is explained for the \Id PID that when both predictor-target correlations are non-zero the minimum edge value occurs at edge value $i$ or $k$. When the synergy moves from zero to a positive value, this means that the minimum edge value has changed from being $k$ to being equal to $i$, and vice-versa.  For a given value of $p$, one can explore the regions in $(q, r)$ space at which such transitions take place. In Fig.~\ref{ZS}, this region of zero synergy is shown, given four different values of $p$. The boundary of each of the regions is where the synergy component changes from positive synergy to zero synergy, or vice-versa. 
 
 The plots in Fig. 5 show that synergy is non-zero (positive) whenever $q$ and $r$ are of opposite sign. When the predictor-predictor correlation, $p$, is 0.05 there is also positive synergy for large regions, defined by $q r -p>0$,  when $q$ and $r$ have the same sign. As $p$ increases the regions of zero synergy change shape, initially increasing in area and then declining as $p$ becomes quite large ($p=0.75$). As $p$ is increased further the zero-synergy bands narrow and so zero synergy will only be found when $q$ and $r$ are close to being equal. 
 
 When $p$ is negative, the corresponding plots are identical to those with positive $p$ but rotated counter clockwise by $\pi/2$ about the point $q=0, r=0.$ Hence, synergy is present when $q$ and $r$ have the same sign. When $q$ and $r$ have opposite signs, there is also positive synergy for regions defined by $q r -p < 0.$
 
 The case of $p=0$ is of interest and there are no non-zero admissible values of $q$ and $r$ (where the covariance matrix is positive definite)  where the synergy  is equal to zero. Hence the system will have synergy in this case unless $q=0$ or $r=0$. This can be seen from the \Id synergy expression in Example 3.
\section{ Multivariate continuous  predictors and target}
We now extend the results developed in Section 2 and consider the case where the three continuous variables $X_0, X_1, Y$ become random vectors $ \mathbf{X}_0, \mathbf{X}_1, \mathbf{Y}$, of dimensions $n_0, n_1, n_2,$ respectively, with mean vectors equal to a zero vector of lengths $n_0, n_1, n_2$, respectively, and covariance matrices equal to an identity matrix of the respective sizes $n_0 \times n_0, n_1 \times n_1, n_2 \times n_2.$ The fact that there is no loss of generality in making these assumptions will be explained in Section 3.4.  We stack these random vectors into the random vector $\mathbf{Z}$, where $\mathbf{Z}$ has dimension $n_0 + n_1 + n_2,$ and assume that $\mathbf{Z}$ has a multivariate Gaussian distribution with p.d.f. f$(\mathbf{x}_0, \mathbf{x}_1, \mathbf{y}),$ mean vector $\mathbf{0}$ and covariance matrix given by
\begin{equation}
\Sigma_Z = \begin{bmatrix}   \id{n_0} & P & Q \\ P^T & \id{n_1} &R \\Q^T & R^T & \id{n_2}   \end{bmatrix}, \label{sig2}
\end{equation}
where the matrices $P, Q, R$ are of size $n_0 \times n_1, n_0 \times n_2, n_1 \times n_2$, respectively, and are the cross-covariance (correlation) matrices between the three pairings of the three vectors $ \mathbf{X}_0, \mathbf{X}_1, \mathbf{Y},$ and so
\begin{equation}
\mathbb{E}(\mathbf{X}_0 \mathbf{X}_1^T) = P, \quad \mathbb{E}(\mathbf{X}_0 \mathbf{Y}^T) = Q, \quad \mathbb{E}(\mathbf{X}_1 \mathbf{Y}^T) = R, \label{cross}
\end{equation}
defined on $\mathbb{R}^m$, where $m = n_0 + n_1 + n_2.$

\subsection{Properties of the matrices P, Q, R, and the inverse matrix of $\Sigma_Z$}
We require some matrix results, which will be proved in Appendix B.

\begin{lemma}
Suppose that a symmetric matrix $M$ is partitioned as
$$ M = \begin{bmatrix} M_{11} & M_{12} & M_{13} \\ M_{12}^T & M_{22} & M_{23} \\ M_{13}^T & M_{23}^T & M_{33} \end{bmatrix}, \label{siginv}$$
where the diagonal blocks $M_{11}, M_{22}, M_{33}$ are symmetric and square. Then if $M$ is positive definite these diagonal blocks are also positive definite, and so nonsingular.

\end{lemma}

\begin{lemma}
When the covariance matrix $\Sigma_Z$ in~$\eqref{sig2}$  is positive definite then the following matrices are also positive definite, and hence nonsingular:
$$ \id{n_1} - P^TP,\,\, \id{n_0} - PP^T, \,\,\id{n_2}-R^TR,\,\, \id{n_1}-RR^T, \,\,\id{n_2}-Q^TQ, \,\,\id{n_0}-QQ^T.$$ Also, the determinant of each of these matrices is positive and bounded above by unity, and it is equal to unity if, and only if, the matrix involved is the zero matrix. Furthermore,
$$ \begin{vmatrix}  \id{n_0} & P \\ P^T & \id{n_1}   \end{vmatrix} = | \id{n_1} - P^TP |. $$

\end{lemma}

With these results in place, we now present the inverse of $\Sigma_Z$, which is equal to the concentration matrix $K$. It was determined by solving simultaneous equations for block matrices and we omit the details. It is
\begin{equation}
K = \Sigma_Z^{-1} = \begin{bmatrix}  A & U & V \\
                                   U^T & B & W \\
                                 V^T & W^T & C \end{bmatrix}, \label{BG8}
\end{equation}
where
\begin{align}
U &=    (\id{n_0} - QQ^T)^{-1}(QR^T -P) B  \label{u} \\
V&=    A (P R-Q)(\id{n_2} -R^TR)^{-1}   \label{v}\\
W&=    (\id{n_1} -P^TP)^{-1}(P^TQ-R)C   \label{w}  \\
A &= \begin{bmatrix} \id{n_0} - PP^T -(PR-Q)(\id{n_2}-R^T R)^{-1}(P R - Q)^T \end{bmatrix}^{-1} \label{a}\\
B &= \begin{bmatrix} \id{n_1} - RR^T -(QR^T-P)^T(\id{n_0}-QQ^T)^{-1}(Q R^T - P) \end{bmatrix}^{-1} \label{b}\\
C &= \begin{bmatrix} \id{n_2} - Q^TQ -(P^TQ-R)^T (\id{n_1}-P^T P)^{-1}(P^T Q - R) \end{bmatrix}^{-1} \label{c}
\end{align}
The various inverses used in~$\eqref{u}$-$\eqref{c}$ are valid for the following reasons. The matrix $\Sigma_Z$ is positive definite, and so its inverse is also positive definite. By Lemma 1, $A, B, C$ are positive definite and so invertible, which means in turn that their inverses are invertible. From Lemma 2, we have that the matrices $\id{n_1} -P^TP, \,\, \id{n_0}-QQ^T,\,\, \id{n_2}-R^TR$ are invertible. Therefore the sub-matrices in the inverse of $\Sigma_Z$ in~$\eqref{u}$-$\eqref{c}$ are well-defined.

\subsection{Block Gaussian graphical models}
As in Section 1.3, we will consider graphical models to express the conditional  independences in the probability distribution for $\mathbf{Z}$, although each graph will still have three vertices, with each vertex representing one of the random vectors, $ \mathbf{X}_0, \mathbf{X}_1, \mathbf{Y}.$ Each graph can be thought of as a block independence graph. This means that only dependences between pairs of vectors will be represented, while there will be no dependences among the variables within each of the three random vectors, since they are mutually independent. The models which express conditional dependences have the same format as in Table~\ref{GM1} in Section 1.3 and we use the same notation again here, the only difference being to express $ \mathbf{X}_0, \mathbf{X}_1, \mathbf{Y}$ in a bold font.
We term these models 'block graphical models' since we are treating each random vector as the block containing a number of  mutually independent random variables. Here is an illustration of such a model:

\begin{figure}[H]
\centering

\begin{tikzpicture}[
 roundnode/.style={circle, draw =black, fill =black, scale =0.2}]
   \draw[blue, very thick] (0,0)rectangle (2,2);
   \node[above] at (1, 2.1) {$\mathbf{X}_0$};

 \node[roundnode, label= {$X_{01}$}, above] at (0.5, 1.3) { };
 \node[roundnode, label= {$X_{02}$}, above] at (1.5, 1.3) { };
  \node[roundnode, label= {$X_{03}$}, above] at (0.5, 0.3) { };
   \node[roundnode, label= {$X_{04}$}, above] at (1.5, 0.3) { };
   \draw[blue, very thick] (4,2)rectangle (6,4);
 \node[above] at (5, 4.1) {$\mathbf{X}_1$};
 \node[roundnode, label= {$X_{11}$}, above] at (4.5, 3.3) { };
 \node[roundnode, label= {$X_{12}$}, above] at (5.5, 3.3) { };
  \node[roundnode, label= {$X_{13}$}, above] at (5, 2.3) { };
  
   \draw[blue, very thick] (8,0)rectangle (10,2);
    \node[above] at (9, 2.1) {$\mathbf{Y}$};
   \node[roundnode, label= {$Y_{1}$}, above] at (8.5, .6) { };
 \node[roundnode, label= {$Y_{2}$}, above] at (9.5, .6) { };
 
 \draw[-, thick] (2,1) -- (4,3);
 \draw[-, thick] (6,3) -- (8,1);
  \end{tikzpicture}
  \caption{An illustration of a block graphical model for the random vectors  $\mathbf{X}_0$, $\mathbf{X}_1$ and $\mathbf{Y}$. $\mathbf{X}_0$ contains four random variables, while $\mathbf{X}_1$ has three and $\mathbf{Y}$ has two. This model expresses the conditional independence of $\mathbf{X}_0$ and $\mathbf{Y}$ given  $\mathbf{X}_1$. In this model, the bivariate marginals $\mathbf{X}_0\mathbf{X}_1$ and $\mathbf{X}_1\mathbf{Y},$ as well as lower-order marginals, are fixed.} \label{blockGM}
  \end{figure}
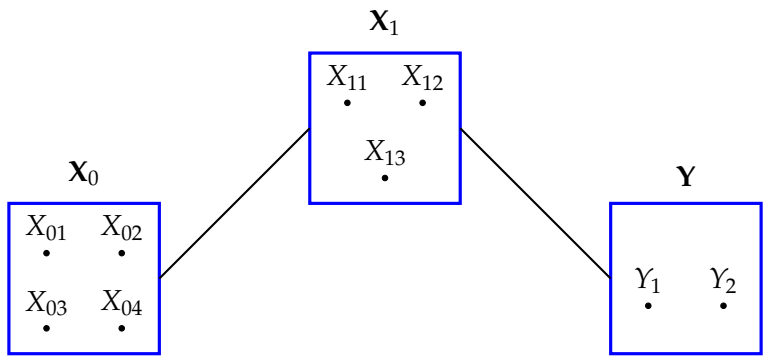
  
\begin{figure}[H]
      \begin{center}
            {\large
\begin{tikzpicture}
  \node (max) at (0,4) {$M_8: \mathbf{\textcolor{blue}{X_0X_1}, \textcolor{mygreen}{X_0Y}, \textcolor{red}{X_1Y}}$};
  \node (a) at (-3,2) {$ M_5: \mathbf{\textcolor{blue}{X_0X_1}, \textcolor{mygreen}{X_0Y}  }$};
  \node (b) at (0,2) {$ M_6: \mathbf{\textcolor{blue}{X_0X_1},  \textcolor{red}{X_1Y}}$};
  \node (c) at (3,2) {$  M_7:  \mathbf{\textcolor{mygreen}{X_0Y}, \textcolor{red}{X_1Y}}$};
  \node (d) at (-3,0) {$ M_2: \mathbf{\textcolor{blue}{X_0X_1}, Y}$};
  \node (e) at (0,0) {$ M_3: \mathbf{\textcolor{mygreen}{X_0Y}, X_1}$};
  \node (f) at (3,0) {$M_4:   \mathbf{\textcolor{red}{X_1Y}, X_0}$};
  \node (min) at (0,-2) {$M_1: \mathbf{X_0, X_1, Y}$};
  \draw[mygreen, thick] (d) -- (a) node[midway, fill=white] {d};
   \draw[mygreen, thick] (min) -- (e) node[midway, fill=white] {b};
    \draw[mygreen, thick] (b) -- (max) node[midway, fill=white] {k};
  \draw [red, thick] (a) -- (max) node[midway, fill=white]{j};
  \draw[red, thick] (min) -- (f)  node[midway, fill=white] {c};
    \draw [red, thick] (d) -- (b) node[midway, fill=white]{f};
      \draw [red, thick] (e) -- (c) node[midway, fill=white]{h};
  \draw[mygreen, thick] (f) -- (c) node[midway, fill=white] {i};
\end{tikzpicture}
}
  \caption{A dependency lattice of block graphical models. Edges coloured green (b, d, i, k) correspond to adding the set of constraints within $\mathbf{X_0Y}$ to the model immediately below. Edges coloured red (c, f, h, j) correspond to adding the set of constraints within $\mathbf{X_1Y}$ to the model immediately below. The two relevant sub-lattices are shown here. \label{BlockGM}}
      \end{center} 
      \end{figure}
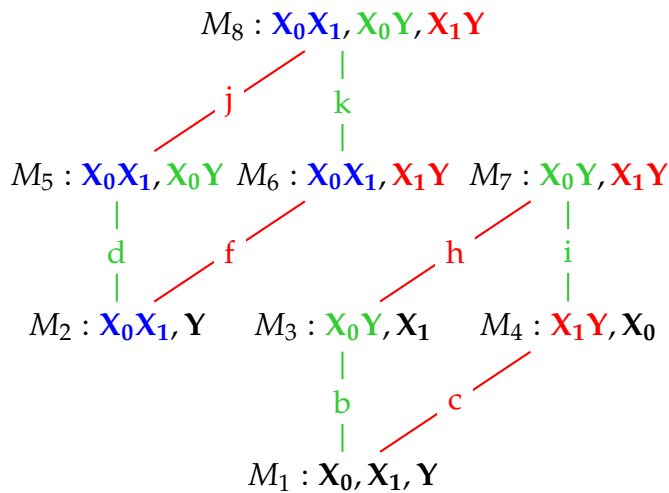

 The model in Fig. 6 is the block version of model $G_6$ from Table~\ref{GM1} and denoted as $\mathbf{X}_0\mathbf{X}_1, \mathbf{X}_1\mathbf{Y}.$ A product term, such as $\mathbf{X}_0\mathbf{X}_1$, encapsulates correlations between each random variable in $\mathbf{X}_0$ and each random variable in $\mathbf{X}_1$. For example, in Fig. 6 there are 12 correlations  between the elements of $\mathbf{X}_0$ and $\mathbf{X}_1$, and 6 correlations between the elements of $\mathbf{X}_1$ and $\mathbf{Y}$. Using the block notation provides some simplicity, for otherwise one would be required to write expressions such as
 $$ \mathbf{X}_1\mathbf{Y} = X_{11}Y_1 + X_{11}Y_2 + X_{12}Y_1 + X_{12}Y_2 +X_{13}Y_1 + X_{13}Y_2$$
for the set of constraints within each  block interaction term. The block graphical models in the multivariate version of the dependency lattice are given in Fig.~\ref{BlockGM}.
 We now define the conditional independence constraints for the block versions of model $G_1 \ldots G_8$ in Table~\ref{GM1} and determine some of their estimated covariance matrices.  The block version of model $G_8$ has no conditional independences. Hence, no block zeroes are imposed on the concentration matrix $K,$ and the estimated covariance matrix for this model is $\hat{\Sigma}_8 = \Sigma_Z$, which means that model $G_8$ is equal to the given distribution of $\mathbf{Z}$.

Consider the block version of model $G_7,$ in Table~\ref{GM1}. In $G_7$, $ \mathbf{X}_0$ and $\mathbf{X}_1$ are conditionally independent  given $\mathbf{Y}$ and so  we apply the constraint $U =0$ in the concentration matrix $K$ in~$\eqref{BG8}$.  From~$\eqref{u}$, this block constraint is
\begin{equation}
(\id{n_0} - QQ^T)^{-1}(QR^T -P) B = 0. \label{cons1}
\end{equation}
Given the results stated at the end of Section 3.1, we can pre-multiply by $\id{n_0}-QQ^T$ and post-multiply by $B^{-1}$ in~$\eqref{cons1}$ to obtain the required block constraint as $QR^T -P =0,$ that is $P = QR^T.$ Hence, the estimated covariance matrix for block model $G_7$ is
\begin{equation}
\hat{\Sigma}_7 = \begin{bmatrix}   \id{n_0} & QR^T & Q \\ RQ^T & \id{n_1} &R \\Q^T & R^T & \id{n_2}   \end{bmatrix}.
\label{BK7}
\end{equation}

Applying a similar argument to the expressions for $V$ and $W$ in~$\eqref{v}$ and~$\eqref{w}$, it can be shown that for the conditions for conditional independence between $\mathbf{X}_0$ and $\mathbf{Y}$ given  $\mathbf{X}_1$ in model $G_6$ the block constraint is $Q =P R$, while for the  conditional independence between $\mathbf{X}_1$ and $\mathbf{Y}$ given  $\mathbf{X}_0$ in model $G_5$  the block constraint is $R = P^TQ.$ Hence, the estimated covariance matrices for block models $G_5$ and $G_6$ are
\begin{equation}
\hat{\Sigma}_5 = \begin{bmatrix}   \id{n_0} & P  & Q \\ P^T & \id{n_1} &P^TQ \\Q^T & Q^P & \id{n_2}   \end{bmatrix}, \quad \text{and} \quad 
\hat{\Sigma}_6 = \begin{bmatrix}   \id{n_0} & P & PR \\ P^T & \id{n_1} &R \\R^TP^T & R^T & \id{n_2}   \end{bmatrix}.  \label{BK56}
\end{equation}
In block model $G_2$ both of the conditional independences defining block models $G_5$ and $G_6$ are present. So, the conditional independence constraints are $R= P^TQ$ and $Q=PR$. Combining them gives $R = P^TP R,$ which may be written as $(\id{n_1} - P^TP) R=0$. By Lemma 2, we may pre-multiply by the inverse of $\id{n_1} - P^TP$ to obtain  $R=0$, which in turn implies that $Q=0$. Hence the conditional independence constraints for block model $G_2$ are $Q=0, R=0$ and so the estimated covariance matrix is
\begin{equation}
\hat{\Sigma}_2 = \begin{bmatrix}   \id{n_0} & P & 0 \\ P^T & \id{n_1} &0 \\0& 0& \id{n_2}   \end{bmatrix}.  \label{BK6}
\end{equation}
The other estimated covariance matrices can be derived in a similar fashion. As in Section 2.1, it is the case that applying the conditional independence constraints also ensures that the required marginal distributions in each of the block graphical models are equal to the corresponding marginal distributions in the given distribution of $\mathbf{Z}$. We also note, in particular, that model $M_2$ in Fig.~\ref{BlockGM} has the same conditional independences as those present in block graphical model $G_2$, and this is true for each of the maximum entropy models $M_i$, and so when finding the form of these models in the next section we apply in each case the conditional independence constraints satisfied by the block graphical model $G_i$.

\subsection{Maximum entropy distributions}
We are given the distribution of $Z$ which is multivariate Gaussian with zero mean vector and covariance matrix $\Sigma_Z$ in~$\eqref{sig2}$, and has p.d.f. $f(\mathbf{z}) \equiv f(x_0, x_1, y).$ For each of the models $M_1 \ldots M_8$, we will determine  the  p.d.f. of the maximum entropy solution $g(\mathbf{z}) \equiv g(x_0, x_1, y)$ subject to the constraints
\begin{equation}
\int_{\mathbb{R}^3} \mathbf{z} \,g(\mathbf{z}) \,\, d\mathbf{z} = \mathbf{0}, \quad \int_{\mathbb{R}^3} g(\mathbf{z}) \,\, d\mathbf{z}  = 1, \quad g(\mathbf{z}) > 0, \label{MEC1}
\end{equation}
and the separate constraint for model $M_i$
\begin{equation}
\int_{\mathbb{R}^3}\mathbf{z z}^T g(\mathbf{z}) \,\, d \mathbf{z} = \hat{\Sigma}_i, \label{MEC2}
\end{equation}
as well as the conditional independence constraints given in Table~\ref{CIcon2}.
 \setlength{\tabcolsep}{1em}
 \begin{table}[H]
\centering
\caption{Conditional independence constraints satisfied by the block Gaussian graphical models $G_1 \ldots G_8$ that are applied when determining the maximum entropy models $M_1 \ldots M_8$.    } \label{CIcon2}
\begin{tabular}{lll} \toprule
$M_1: P = QR^T, Q=PR, R=P^TQ$& & \\
$M_2: Q=PR, R=P^TQ$&  $M_3: P=QR^T, R=P^TQ $& $M_4: P=QR^T,  Q=PR$ \\
$M_5: R=P^TQ $&  $M_6:  Q=PR$& $M_7: P = QR^T$ \\

\bottomrule
\end{tabular}
\end{table}
For model $M_8$, the estimated covariance matrix $\hat{\Sigma}_8 = \Sigma_Z$ and so the maximum entropy distribution $M_8$ is equal to block graphical model $G_8$, which is equal to the  given distribution of $\mathbf{Z}$. Similarly, the maximum entropy model $M_7$ is equal to the block graphical model $G_7$, which is multivariate Gaussian with zero mean vector and covariance matrix $\hat{\Sigma}_7,$ defined in~$\eqref{BK7}$, and so on. Hence we can state our results in Proposition 3.
\begin{proposition}
The distributions of maximum entropy, $M_1 \ldots M_8$, subject to the constraints~$\eqref{MEC1}$-$\eqref{MEC2}$ and the conditional independence constraints in Table~\ref{CIcon2}
are block  Gaussian  graphical models $G_1 \ldots G_8$ having mean vector $\mathbf{0}$ and with the covariance matrices $\hat{\Sigma}_i, \,\,(i=1, \ldots 8),$  given below in Table~\ref{MEcov2}
\end{proposition}
\begin{table}[H]
\centering
\caption {Covariance matrices for the Gaussian block graphical models in Proposition 3. } \label{MEcov2}
\begin{tabular}{lclc} \toprule
Model & $\hat{\Sigma}_i$ & Model & $\hat{\Sigma}_i$ \\ \midrule
$M_1: \mathbf{X}_0, \mathbf{X}_1, \mathbf{Y}$ & $\begin{bmatrix} \id{n_0} & 0& 0\\ 0 & \id{n_1} & 0\\ 0& 0& \id{n_2} \end{bmatrix} $ &$M_5: \mathbf{X}_0\mathbf{X}_1, \mathbf{X}_0\mathbf{Y}$    &   $\begin{bmatrix} \id{n_0} & P & Q\\ P^T & \id{n_1} & P^TQ\\ Q^T& Q^TP& \id{n_2} \end{bmatrix} $ \\\\
$M_2: \mathbf{X}_0\mathbf{X}_1, \mathbf{Y}$ & $\begin{bmatrix} \id{n_0} & P & 0\\ P^T & \id{n_1} & 0\\ 0& 0& \id{n_2} \end{bmatrix} $ & $M_6: \mathbf{X}_0\mathbf{X}_1, \mathbf{X}_1\mathbf{Y}$   &   $\begin{bmatrix} \id{n_0} & P & PR\\ P^T & \id{n_1} & R\\ R^TP^T& R^T& \id{n_2} \end{bmatrix} $   \\\\
$M_3: \mathbf{X}_0\mathbf{Y}, \mathbf{X}_1$ & $\begin{bmatrix} \id{n_0} & 0 & Q\\ 0 & \id{n_1} & 0\\ Q^T& 0& \id{n_2} \end{bmatrix} $ &$M_7:\mathbf{X}_0\mathbf{Y}, \mathbf{X}_1\mathbf{Y}$     &   $\begin{bmatrix} \id{n_0} & QR^T & Q\\  RQ^T & \id{n_1} & R\\ Q^T& R^T& \id{n_2} \end{bmatrix} $ \\\\
$M_4: \mathbf{X}_1\mathbf{Y}, \mathbf{X}_0$ & $\begin{bmatrix} \id{n_0} & 0 & 0\\ 0 & \id{n_1} & R\\ 0& R^T& \id{n_2} \end{bmatrix} $ & $M_8:\mathbf{X}_0\mathbf{X}_1, \mathbf{X}_0\mathbf{Y}, \mathbf{X}_1\mathbf{Y}$   & $\begin{bmatrix} \id{n_0} & P& Q\\ P^T & \id{n_1} & R\\ Q^T& R^T& \id{n_2} \end{bmatrix} $   \\

\bottomrule
\end{tabular}
\end{table}
We can now check by inspecting the $\hat{\Sigma}_i$ entries in Table~\ref{MEcov2} that particular marginal distributions involving two blocks, such as $\mathbf{X}_1\mathbf{Y}$, are  the same in all of the models and also equal to the marginal distribution in the given distribution.  For example, the block interaction term $\mathbf{X}_1\mathbf{Y}$ is present in models $M_4, M_6, M_7, M_8$. The distribution of $[\mathbf{X}_1 \,\,\, \mathbf{Y}]^T$ is multivariate normal with mean vector equal to a zero vector  and covariance matrix given by the bottom right 2 by 2 block matrix in $\hat{\Sigma}_Z$~\cite[p. 63]{MKB}. For each of  these four models, we can see by inspection that this covariance matrix is $\begin{bmatrix} I & R \\ R^T & I \end{bmatrix},$ and so the ($\mathbf{X}_1, \mathbf{Y})$ marginal distribution is the same in all of these four models in which this particular block interaction term has been fitted. Since $M_8$ is equal to the given distribution of $\mathbf{Z}$, it follows that this marginal distribution is the same as in the given distribution. Similar checks can be made regarding the other marginal distributions involving two blocks to find that a similar conclusion applies also to them. We can also check that the single-block terms, such as $\mathbf{X}_1$ have the same distribution. This term has been fitted in all eight models. Since the mean vector of $\mathbf{X}_1$ is a zero vector, its distribution is determined by its covariance matrix. This is given by the central [2, 2] sub-matrix in $\hat{\Sigma}_Z$. Inspection of the fitted $\hat{\Sigma}_Z$ covariance matrices in Table~\ref{MEcov2} reveals that the relevant matrix $\id{n_1}$ is the same in all eight models. Hence this block marginal distribution is fixed in the eight maximum entropy distributions.

\subsection{Mutual information}
It was claimed in Section 3 that there is no loss of generality in assuming that the mean vectors of $\mathbf{X}_0, \mathbf{X}_1, \mathbf{Y}$ are a zero vector and that their covariance matrices are an identity matrix, of the required sizes. We will now demonstrate this, by calculating the mutual information $I(\mathbf{X}_0, \mathbf{X}_1; \mathbf{Y}),$ using the general form of covariance matrix (which is partitioned conformably to $\Sigma_Z$ in~$\eqref{sig2}$):
\begin{equation}
\Sigma = \begin{bmatrix} \Sigma_{00} & \Sigma_{01} & \Sigma_{02} \\\Sigma_{01}^T & \Sigma_{11} & \Sigma_{12} \\
 \Sigma_{02}^T & \Sigma_{12} ^T& \Sigma_{22}  \end{bmatrix} =   \begin{bmatrix} \Sigma_{00}^{\tfrac{1}{2}} & 0& 0 \\ 0 & \Sigma_{11}^{\tfrac{1}{2}} & 0 \\
 0&0 & \Sigma_{22}^{\tfrac{1}{2}}  \end{bmatrix}^T \begin{bmatrix} \id{n_0} & P & Q \\ P^T &\id{n_1} &R \\
 Q^T & R ^T& \id{n_2}  \end{bmatrix}  \begin{bmatrix} \Sigma_{00}^{\tfrac{1}{2}} & 0 & 0 \\0  & \Sigma_{11}^{\tfrac{1}{2}} & 0 \\
0 & 0& \Sigma_{22}^{\tfrac{1}{2}}  \end{bmatrix}, \label{fact}
\end{equation}
where
\begin{equation} P =  \Sigma_{00}^{-\tfrac{1}{2}}  \Sigma_{01} \Sigma_{11}^{-\tfrac{1}{2}}, \quad Q = \Sigma_{00}^{-\tfrac{1}{2}} \Sigma_{02} \Sigma_{22}^{-\tfrac{1}{2}}, \quad R = \Sigma_{11}^{-\tfrac{1}{2}} \Sigma_{12} \Sigma_{22}^{-\tfrac{1}{2}}. \label{trans} \end{equation}
 Since $\Sigma_{ii}$ (i=0, 1, 2) is positive definite (by Lemma 1) it has a positive definite square root $\Sigma_{ii}^{\tfrac{1}{2}}$~\cite[pp. 405-6]{HJ}. Therefore, using standard properties of determinants,
\begin{equation}
|\Sigma | = |\Sigma_{00}||\Sigma_{11}| |\Sigma_{22}| \begin{vmatrix} \id{n_0} & P & Q \\ P^T & \id{n_1}&R \\
 Q^T & R^T& \id{n_2}  \end{vmatrix}. \label{detprod}
\end{equation}
From~\cite{MKB}, we can state the following marginal distributions.
$$
\mathbf{X}_0 \sim N(\mathbf{0}, \Sigma_{00}), \quad \mathbf{X}_1 \sim N(\mathbf{0}, \Sigma_{11}), \quad  \mathbf{Y} \sim N(\mathbf{0}, \Sigma_{22}), $$
and 
\begin{equation}[\mathbf{X}_0 \,\, \mathbf{X}_1]^T, \quad [\mathbf{X}_0 \,\, \mathbf{Y}]^T, \quad [\mathbf{X}_1 \,\, \mathbf{Y}]^T \label{marg1}
\end{equation}
are multivariate normal with covariance matrices
\begin{equation} \begin{bmatrix} \Sigma_{00} & \Sigma_{01} \\ \Sigma_{01}^T & \Sigma_{11} \end{bmatrix}, \quad  \begin{bmatrix} \Sigma_{00} & \Sigma_{02} \\ \Sigma_{02}^T & \Sigma_{22} \end{bmatrix}, \quad  \begin{bmatrix} \Sigma_{11} & \Sigma_{12} \\ \Sigma_{12}^T & \Sigma_{22} \end{bmatrix},  \label{marg2}\end{equation}
respectively.
The formula for the entropy of a multivariate Gaussian distribution is required. For a $k$-dimensional random variable $\mathbf{W}$ following a Gaussian distribution with mean vector $\boldmath{\mu}$ and covariance matrix $\Sigma$, the  entropy in the distribution of $\boldmath{W}$ is~\cite{CT}
\begin{equation}
H(\mathbf{W}) = \frac{k}{2} + \frac{k}{2} \log(2 \pi e) + \frac{1}{2} \log |\Sigma|, \label{MVNent}
\end{equation}

Using the formula for entropy in~$\eqref{MVNent}$, and using a similar argument to that which produced~$\eqref{detprod}$, we may write
\begin{align}
H(\mathbf{Y}) & = \frac{n_2}{2} + \frac{n_2}{2} \log(2 \pi e) + \frac{1}{2} \log |\Sigma_{22}|,\\
H(\mathbf{X}_0, \mathbf{X}_1) & = \frac{(n_0 + n_1)}{2} + \frac{(n_0 +n_1)}{2} \log(2 \pi e) + \frac{1}{2} \log (|\Sigma_{00}||\Sigma_{11}|) +  \frac{1}{2} \log \begin{vmatrix} \id{n_0} & P \\P^T & \id{n_1}   \end{vmatrix}\\
H(\mathbf{X}_0, \mathbf{X}_1, \mathbf{Y}) & = \frac{m}{2} + \frac{m}{2} \log(2 \pi e) + \frac{1}{2} \log (|\Sigma_{00}||\Sigma_{11}||\Sigma_{22}|) + \frac{1}{2} \log \begin{vmatrix} \id{n_0} & P & Q\\ P^T &\id{n_1} &R \\
 Q^T & R ^T& \id{n_2}  \end{vmatrix} 
\end{align}
Therefore, applying a version of~$\eqref{detsig}$ with vector arguments, and using~$\eqref{sig2}$, the total mutual information is given by
\begin{equation}
I(\mathbf{X}_0, \mathbf{X}_1; \mathbf{Y}) = \dfrac{1}{2} \log \begin{vmatrix} \id{n_1} - P^TP  \end{vmatrix} -  \frac{1}{2} \log \begin{vmatrix} \id{n_0} & P & Q\\ P^T &\id{n_1} & R\\
 Q^T & R^T& \id{n_2}  \end{vmatrix} = \frac{1}{2} \log \frac{\begin{vmatrix} \id{n_1}- P^TP  \end{vmatrix}}{|\Sigma_Z|},
\end{equation}
using the fact that (Lemma 2)
$$ \begin{vmatrix} \id{n_0} & P \\ P^T &\id{n_1}  \end{vmatrix} = \begin{vmatrix} \id{n_1} - P^TP  \end{vmatrix}.$$
Therefore, we have demonstrated that the mutual information between the predictors and the target does not depend on the either the mean vectors or the covariance matrices of the individual random vectors $\mathbf{X}_0$, $\mathbf{X}_1$ and $\mathbf{Y}.$
Using~$\eqref{MVNent}$, the distributional results~$\eqref{marg1}$-$\eqref{marg2}$ and similar arguments to that leading to~$\eqref{detprod}$, we state formulae for the other required mutual informations.
 
\begin{align}
I(\mathbf{X}_0; \mathbf{Y}) &= \frac{1}{2} \log  \frac{1}{ \begin{vmatrix} \id{n_2}- Q^TQ  \end{vmatrix}},  \label{mvMI1}\\
I(\mathbf{X}_1; \mathbf{Y}) &= \frac{1}{2} \log  \frac{1}{ \begin{vmatrix} \id{n_2} - R^TR  \end{vmatrix}}, \\
I(\mathbf{X}_0; \mathbf{Y}| \mathbf{X}_1) &= \frac{1}{2} \log \frac{\begin{vmatrix} \id{n_1}- P^TP  \end{vmatrix} \begin{vmatrix} \id{n_2} - R^TR  \end{vmatrix}}{|\Sigma_Z|},  \label{mcondMI1}\\
I(\mathbf{X}_1; \mathbf{Y}|\mathbf{X}_0) &=    \frac{1}{2} \log \frac{\begin{vmatrix} \id{n_1} - P^TP  \end{vmatrix} \begin{vmatrix} \id{n_2} - Q^TQ  \end{vmatrix}}{|\Sigma_Z|} \label{mvMI4}
\end{align}
Expressions for the total mutual information, $I(\mathbf{X}_0, \mathbf{X}_1; \mathbf{Y})$, for models $M_1 \ldots M_8$  are provided in Table~\ref{mitot2}.

\begin{table}[H]
\centering
\caption{Expressions for the predictors-target mutual information for the eight models in the dependency lattice in Fig.~\ref{BlockGM}, that are stated   in Table~\ref{MEcov2} \label{mitot2}}
\begin{tabular}{ll} \toprule
$M_8:   \frac{1}{2} \log \frac{\begin{vmatrix} \id{n_1} - P^TP  \end{vmatrix}}{\begin{vmatrix} \Sigma_Z \end{vmatrix}}$& $M_4: I(\mathbf{X}_1; \mathbf{Y})$\\\\
$M_7:     \frac{1}{2} \log \frac{\begin{vmatrix} \id{n_1} - RQ^TQR^T  \end{vmatrix}}{\begin{vmatrix} \id{n_2} - Q^TQ  \end{vmatrix}  \begin{vmatrix} \id{n_2} - R^TR \end{vmatrix}}  $&$M_3: I(\mathbf{X}_0; \mathbf{Y}) $ \\\\
$M_6:  I(\mathbf{X}_1;\mathbf{Y}) = \frac{1}{2} \log \frac{ \scalebox{1.05}{1} }{ \begin{vmatrix} \id{n_2} - R^TR  \end{vmatrix}} $ & $M_2: 0$\\\\
$M_5: I(\mathbf{X}_0; \mathbf{Y}) =  \frac{1}{2} \log \frac{ \scalebox{1.05}{1} }{ \begin{vmatrix} \id{n_2} - Q^TQ  \end{vmatrix}} $&$M_1: 0 $\\

\bottomrule
\end{tabular}
\end{table}

\subsection{The \Id PID for multivariate Gaussian predictors and targets}

Using the expressions for the total mutual information between predictors and the target in Table~\ref{mitot2},  formulae for the edge values that are used in the construction of the \Id PID, are given in Table~\ref{ev2}. They are computed by subtracting the mutual informations of the relevant models in Fig. 7; for example, $k$ is computing by subtracting the mutual information in model $M_6$ from that in model $M_8$.
 \setlength{\tabcolsep}{2em}
\begin{table}[H]
\centering
\caption{Expression for the edge values in the dependency lattice in Fig.~\ref{BlockGM} that are used to determine the unique informations. } \label{ev2}
\begin{tabular}{ll} \toprule \\
$ b = d = I(\mathbf{X}_0; \mathbf{Y}) =   \frac{1}{2} \log  \frac{ \scalebox{1.05}{1} }{ \begin{vmatrix} \id{n_2} - Q^TQ  \end{vmatrix}} $ &  $ c = f = I(\mathbf{X}_1; \mathbf{Y}) =  \frac{1}{2} \log \frac{ \scalebox{1.05}{1} }{ \begin{vmatrix} \id{n_2} - R^TR  \end{vmatrix}} $ \\\\
$i =  \frac{1}{2} \log \frac{\begin{vmatrix} \id{n_1}- RQ^TQR^T  \end{vmatrix}}{\begin{vmatrix} \id{n_2} - Q^TQ  \end{vmatrix}  \begin{vmatrix} \id{n_2} - R^TR \end{vmatrix}}  - I(\mathbf{X}_1; \mathbf{Y})   $& $ h = \frac{1}{2} \log \frac{\begin{vmatrix} \id{n_1} - RQ^TQR^T  \end{vmatrix}}{\begin{vmatrix} \id{n_2}- Q^TQ  \end{vmatrix}  \begin{vmatrix} \id{n_2} - R^TR \end{vmatrix}}  - I(\mathbf{X}_0; \mathbf{Y})    $\\\\
$ k =    \frac{1}{2} \log \frac{\begin{vmatrix} \id{n_1}- P^TP  \end{vmatrix}}{\begin{vmatrix} \Sigma_Z \end{vmatrix}}   - I(\mathbf{X}_1; \mathbf{Y})    $&$ j =      \frac{1}{2} \log \frac{\begin{vmatrix} \id{n_1} - P^TP  \end{vmatrix}}{\begin{vmatrix} \Sigma_Z \end{vmatrix}} -I(\mathbf{X}_0; \mathbf{Y})     $\\\\

\bottomrule
\end{tabular}
\end{table}
Given the edge values in Table~\ref{ev2}, we can form the \Id PID for multivariate Gaussian predictors and targets.
\begin{align}
\text{unq0} &= \min\{b, d, i, k\},  &  \text{red} &= I(\mathbf{X}_0;\mathbf{Y}) - \text{unq0}, \label{mid1}\\
 \text{unq1} &= I(\mathbf{X}_1;\mathbf{Y}) - \text{red},   & \text{syn} &= I(\mathbf{X}_0;\mathbf{Y}|\mathbf{X}_1) - \text{unq0}. \label{mid2}
\end{align}
We now state some results for this PID. Proofs are given in Appendix C.

\begin{proposition} 

For two multivariate Gaussian predictors, $\mathbf{X}_0, \mathbf{X}_1$,  and one multivariate Gaussian target, $\mathbf{Y}$, the PID defined in Table~\ref{ev2} and~$\eqref{mid1}$-$\eqref{mid2}$ has the following properties.
\begin{itemize}
\item[(a)] This \Id PID possesses  consistency as well as the core axioms of non-negativity,  self-redundancy, monotonicity, symmetry and identity.
\item[(b)] When unq0 is equal to $b$ or $d$, the the redundancy component is zero.
\item[(c)] When unq0 is equal to $i$, the redundancy and both unique informations are constant with respect to the correlation matrix $P$  between the two predictors, $\mathbf{X}_0, \mathbf{X}_1$.
\item[(d)] When neither predictor and the target are  independent, then unq0 is equal to either $i$ or to $k$.

\item[(e)] When unq0 is equal to $k$, the synergy component is zero.
\item[(f)] The redundancy component in the \Imm PID is greater than or equal to the redundancy component in the \Id PID with equality if, and only, if at least one of the following conditions holds: (i) either predictor and the target are independent; (ii) either predictor is conditionally independent of the target given the other predictor.
\item[(g)] The synergy component in the \Imm PID is greater than or equal to the synergy  component in the \Id PID with equality if, and only, if at least one of the following conditions holds: (i) either predictor and the  target are independent; (ii) either predictor is conditionally independent of the target given the other predictor.
\item[(h)] The \Id and \Imm PIDs are identical when either $\mathbf{X}_0$ and $\mathbf{Y}$ are conditionally independent given $\mathbf{X}_1$ or  $\mathbf{X}_1$ and $\mathbf{Y}$ are conditionally independent given $\mathbf{X}_0$, and in particular they are identical for models $M_1 \ldots M_6$.  In model $M_7$  the synergy component of \Id is zero.

\end{itemize}

\end{proposition}
 
 \subsection{Examples and Illustrations}
  The multivariate version of the \Id PID was implemented using the edge coefficients in Table~\ref{ev2} together with the PID rules in~$\eqref{mid1}$-$\eqref{mid2}$.
 The  matrices, $P, Q, R$, were given an equi-correlation structure in which all the entries were equal within each matrix:
 \begin{equation}
 P = p \mathbf{1}_{n_0}\mathbf{1}_{n_1}^T, \quad Q = q \mathbf{1}_{n_0}\mathbf{1}_{n_2}^T, \quad R = r \mathbf{1}_{n_1}\mathbf{1}_{n_2}^T,
 \end{equation}
 where $p, q, r$ denote here the constant correlations with each matrix and $\mathbf{1}_n$ denotes an n-dimensional vector whose entries are each equal to unity. 

 Taking $p=0.1, q=0.2, q=0.3, n_0 =4, n_1 =3, n_0=2$, respectively, the covariance (correlation) matrix $\Sigma_Z$ was computed and plots produced of the PIDs as displayed below. The covariance matrix is positive definite only for limited ranges of $p, q, r$. The \Imm PID was computed using the formulae in Section 2.5, but replacing $X_0, X_1, Y$ by their vector counterparts $\mathbf{X}_0, \mathbf{X}_1, \mathbf{Y}$, respectively.

  \setlength{\tabcolsep}{0.5em}
\begin{figure}[H]
\centering

\begin{tabular}{cc}
\begin{subfigure}{0.4\textwidth}\centering\includegraphics[scale=0.5]{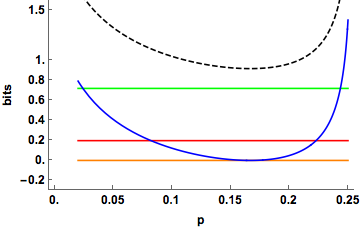}
\caption{$ \text{\Imm}, q = 0.3, r =0.2$} \end{subfigure} &\begin{subfigure}{0.4\textwidth}\centering\includegraphics[scale=0.5]{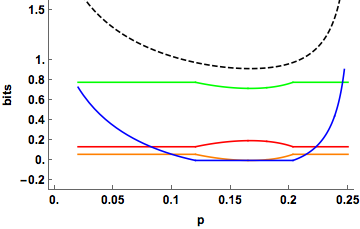}
\caption{$ \text{\Id}, q = 0.3, r= 0.2$} \end{subfigure} \\\\
\end{tabular}
\begin{center}
\includegraphics[scale=0.5]{IdepLegend.png} 
\end{center}
\begin{tabular}{cc}
\begin{subfigure}{0.4\textwidth}\centering\includegraphics[scale=0.5]{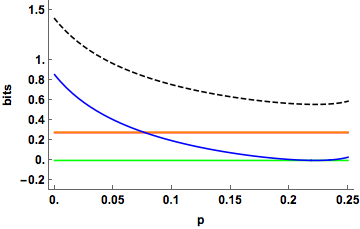}
\caption{$ \text{\Imm}, q =0.2, r = 0.3$}  \end{subfigure}&
\begin{subfigure}{0.4
\textwidth}\centering\includegraphics[scale=0.5]{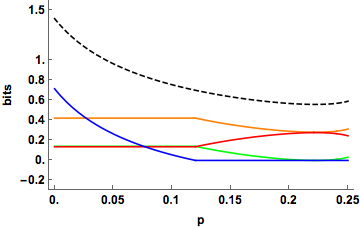}
\caption{$ \text{\Id}, q = 0.2, r = 0.3$} \end{subfigure}
\end{tabular}
 \caption{The \Imm and \Id PID components are plotted for a range of values of the correlation ($p$) between the two predictors. Two combinations of the correlations ($q, r$) between each predictor and the target are displayed. The total mutual information $I(X_0, X_1; Y)$ is also shown as a dashed black curve.   }
\end{figure}

Fig. 8 shows some plots of the multivariate \Imm and \Id PIDs as a function of $p$,  for particular values of $q$ and $r$. These plots display similar characteristics to those shown in Fig. 3, Section 2.7. Some further plots are displayed in Fig.9. This time the PIDs are shown for increasing values of $q (=r)$, for two values of $p$. Again, these plots have similar characteristics to those considered in Fig. 4, Section 2.7.
\begin{figure}[H]
\centering

\begin{tabular}{cc}
\begin{subfigure}{0.4\textwidth}\includegraphics[scale=0.5]{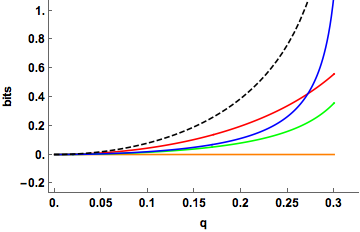}
\caption{$ \text{\Imm}, p=0.1$} \end{subfigure} &\begin{subfigure}{0.4\textwidth}\includegraphics[scale=0.5]{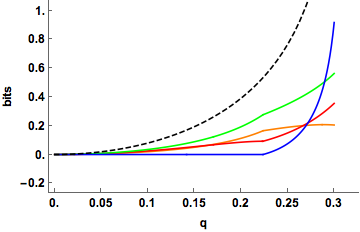}
\caption{$ \text{\Id}, p=0.1$} \end{subfigure} \\\\
\end{tabular}
\begin{center}
\includegraphics[scale=0.5]{IdepLegend.png} 
\end{center}
\begin{tabular}{cc}

\begin{subfigure}{0.4\textwidth}\includegraphics[scale=0.5]{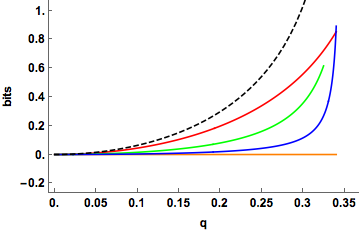}
\caption{$ \text{\Imm},p=0.2$}  \end{subfigure}&
\begin{subfigure}{0.4
\textwidth}\includegraphics[scale=0.5]{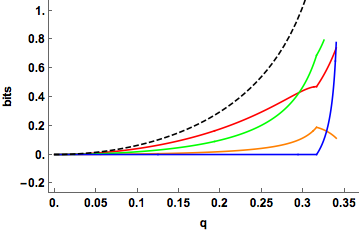}
\caption{$ \text{\Id}, p=0.2$} \end{subfigure}
\end{tabular}
 \caption{The \Imm and \Id PID components are plotted for a range of values of the correlation ($q$) between the predictor $\mathbf{X}_0$ and the target $\mathbf{Y}$. Two combinations of the correlations ($q, r$) between each predictor and the target are displayed. The total mutual information $I(X_0, X_1; Y)$ is also shown as a dashed black curve.   }
\end{figure}

Values of $n_0, n_1, n_2, p, q, r$ were chosen, ensuring that the covariance matrix was positive definite, and  the equi-correlation structure defined in (49) was used. The PID results are presented in the following table.  
\begin{center}
 \begin{tabular}{ccccccc} \toprule
 $(n_0, n_1, n_2)$ & $(p, q, r)$ & PID & unq0 & unq1 & red & syn \\ \midrule
 $(3,4,3)$ & $(-0.15, 0.15, 0.15)$ & \Id & 0.1227 & 0.1865& 0.0406 & 2.4772 \\
 & & \Imm & 0 & 0.0638 & 0.1632 & 2.6000 \\ \midrule
 $(4,4,2)$ & $(-0.2, -0.2, 0.3)$ & \Id & 0.0893 & 0.7293 & 0.1889 & 0.0087 \\
 & & \Imm & 0 & 0.6401 & 0.2782 & 0.0980 \\  \midrule
 $(4,2,4)$ & $(-0.1, 0.15, -0.2)$ & \Id & 0.2336 & 0.1899 & 0.0883 & 0.0345 \\
 & & \Imm & 0.0437 & 0 & 0.2782 & 0.2234 \\
 \bottomrule
 \end{tabular}
 \end{center}
We see rather different compositions of PID components in the three examples as well as some differences between the two methods. For the first system, both methods have a very large value for the synergy component, with \Id having larger values for the unique informations than \Imm but lower redundancy. The two methods produce fairly similar PIDs for the second system, although there are some differences of about 0.09 bit in all of the components. The third system has strong differences between the \Id and \Imm PIDs. \Id has large values for the two unique components along with small values for redundancy and synergy, whereas \Imm has large values for redundancy and synergy and very small values for the uniques.

In these examples, the dimensions of the predictors and target have an impact on the resulting PIDs as well as the correlations.

\begin{example}
Prediction of calcium contents
\end{example}
The multivariate \Id PID and the \Imm PID were applied using data~\cite[p. 145]{AKL} on 73 women involving one set of predictors $\mathbf{X}_0$ (Age, Weight, Height), another set of two predictors $\mathbf{X}_1$ (Diameter of os calcis, Diameter of radius and ulna), and target $\mathbf{Y}$ (Calcium content of heel and forearm). The following results were obtained.
\begin{center}
\begin{tabular}{ccccc} \toprule
PID & unq0 & unq1 & red & syn \\ \midrule
\Id & 0.4077 & 0.0800 & 0.0232 & 0.1408 \\
\Imm & 0.3277 & 0 & 0.1032 & 0.2209 \\
\bottomrule
\end{tabular}
\end{center}
Both PIDs indicate the presence of synergy and a large component of unique information due to the variables in $\mathbf{X}_0$. The \Id PID shows more unique information but less redundancy and synergy than \Imm. To explore matters further, deviance tests~\cite[p. 185]{JW} were performed which compared models $M_1\ldots M_7$ against the saturated model $M_8$. In all seven tests the $p$-values are very small indeed ($ p < 10^{-20}$), indicating that there is very strong evidence to reject models $M_1\ldots M_7$ and that model $M_8$ provides the best explanation of the data. The test of $M_6$ against $M_8$ provided very strong evidence in favour including the term $\mathbf{X}\mathbf{Y}$ in model $M_8$, and so there is little surprise that both PIDs have a large value for unq0. The test of model $M_5$ against model $M_8$ gave very strong evidence that  $\mathbf{X}_1\mathbf{Y}$  should also be included in   model $M_8$.  On this occasion, this has not led to a large value for unq1, so perhaps these two  terms combine to produce synergy and redundancy.

The PIDs were also computed with the same $\mathbf{X}_0$ and $\mathbf{Y}$ but taking $\mathbf{X}_1$ to be another set of four predictors (Surface area, Strength of forearm, Strength of leg, Area of os calcis). The following results were obtained.
\begin{center}
\begin{tabular}{ccccc} \toprule
PID & unq0 & unq1 & red & syn \\ \midrule
\Id & 0.3708 & 0.0186 & 0.0601 & 0 \\
\Imm & 0.3522 & 0 & 0.0787 & 0.0186 \\
\bottomrule
\end{tabular}
\end{center}
In this case, the \Imm and \Id PIDs are very similar, with the main component being due to unique information due to the variables in $\mathbf{X}_0$. The \Id PID indicates zero synergy and almost zero unique information due to the variables in  $\mathbf{X}_1$
Again, deviance tests were performed. In six of the seven tests the approximate $p$-values are very small indeed (less than $3 \times 10^{-5}$). The exception is model $M_5$ for which the deviance test has an approximate $p$-value of 0.98, indicating that model $M_5$ provides the simplest explanation for the data and an extremely good fit to the data. In model $M_5$, it is expected that there will be zero synergy as well as a zero unique component due to the variables in $\mathbf{X}_1$, and this matches quite well the information produced in the PIDs.

When working with real or simulated data it is important to use the correct correlation matrix. In order to use the results given in Table~\ref{ev2} and~$\eqref{mid1}$-$\eqref{mid2}$ it is essential that the input covariance matrix has the structure of $\Sigma_Z$, as given in~$\eqref{sig2}$. The computational approach used here is described in Appendix D.

\section{Discussion}
We have applied the \Id method to obtain bivariate partial information decompositions for Gaussian systems with both univariate and multivariate predictors and targets. We give closed form solutions for all PID terms in both these cases, to allow easy computation of a PID from a covariance or correlation matrix. 
The main properties enjoyed by \Id for Gaussian systems  are the same as those defined in~\cite{JEC}. The characteristics of the \Id PIDs for Gaussian system have been illustrated by graphical exploration as well as numerical examples. 

Given that the \Id method employs a lattice of probability models, Gaussian graphical models, it seems natural when attempting to understand the form of a particular PID to consider formal statistical tests in order to determine which of the models in the lattice best fits the data. Therefore, deviance tests have been used for this purpose. They provide a useful complementary approach, as demonstrated in the examples considered.  

There are now three approaches to the PID for Gaussian systems, \Imm \cite{barrett}, $I_\text{ccs}$ \cite{RI1} and $I_\text{dep}$ \cite{JEC} as developed here. 
While they may agree in some cases, these methods are in general all distinct. 
For \Id and $I_\text{ccs}$ the redundancy and unique information values are not invariant to the predictor-predictor marginal distribution (here $p$ or $P$), and so they are not equivalent to \Imm.
Here, we proved that redundancy and synergy measured with \Imm are never less than those measured with \Id, and are equal in specific circumstances regarding marginal and conditional independence (see Propositions 2 \& 4 (f, g)). 
We note that if the full system or data matches any one of the models $U_1 \ldots U_6$ (or $M_1 \ldots M_6$) then a conditional independence condition is met. This forces one unique component and the synergy component to be equal to zero, and \Imm and \Id are identical.
Therefore, in practice, if any of these models provide an acceptable fit to the data, then the \Id and \Imm  PIDs are likely to be quantitatively very similar.  
If models $U_7, U_8$ (or $M_7, M_8$) provide a better fit to the data, then all four components can be non-zero. By considering the perspective of multivariate linear regression based on the conditional distribution of $\mathbf{Y}$ given $\mathbf{X}_0$ and $\mathbf{X}_1$, as in~\cite{olbrich} for the univariate case, synergy is expected to be present in model $M_8$, and \Id and \Imm can diverge. This is also the case with model $M_7$, although here the synergy component of \Id is zero.
A more thorough comparison of the behaviour of these different measures across families of Gaussian systems could help to illustrate their different interpretations and perhaps shed light on the different approaches to the PID in the discrete case.

As noted, while the \Imm PID has the property that the redundancy component does not depend on the correlation between the predictors this is not true in general for the \Id PIDs.
When there is positive synergy in the \Id PIDs it is the case that redundancy and unique information are invariant to predictor-predictor dependence ($p$), but when synergy is zero this does not hold (see e.g. Figure 3).  
However, as shown, the \Id redundancy and synergy terms are always less than or equal to the corresponding \Imm terms. 
From considering the arguments related to the best fitting models, it seems that in some cases the \Imm approach may overstate redundancy.
Further, for \Imm it is by definition not possible for two predictors to both carry unique information. 
Considering the properties of Gaussian systems and simple noisy additive linear systems this seems unintuitive: if the predictors are independent or anti-correlated but with fixed correlation with the target it seems more natural that, across samples, they each provide a positive unique information contribution to an estimate of the target (see e.g. Example 4). 
Similarly, it also seems intuitive that in a Gaussian setting the amount of information shared between two predictors with fixed target correlation should increase as the correlation between the predictors increases (Figure 3d). 
Further, one would imagine, in general that it should be possible for two variables to carry the same amount of information, but for that information to be different.

Both of these considerations suggest that the dependence on the predictor-predictor marginals in both $I_\text{dep}$ and $I_\text{ccs}$ seems to be more natural for Gaussian systems. 
The invariance to the predictor-predictor marginals was a foundational assumption in the derivation of the method presented in \cite{BROJA}, and was based on a decision theoretic operationalization of unique information.
However, a game theoretic extension of this approach in \cite{RI1} suggests that this invariance is not a natural requirement for a measure of shared information. 
In addition, for Gaussian systems there are existing classical variance based approaches to the problem, such as commonality analysis \cite{SM, muk}, based on semi-partial correlation, or path analysis \cite{MS}, which could provide another perspective on the problem. 
Systematically comparing these methods is an interesting area for future work.

The \Id PIDs presented provide a non-negative decomposition of a joint predictor-target mutual information for Gaussian systems. 
This could have broad applications, from an exploratory statistical tool, to analysis of complex systems and networks. 
Gaussian models or approximations have been used for computing information-theoretic statistics from experimental data in neuroscience and neuroimaging~\cite{MWSLP, InceHBM}.  For example, in neuroimaging there are often statistical effects of a stimulus observed in multiple recorded responses (for example different brain regions, or different temporal offsets from stimulation). Methods such as the PID can provide a practical tool to relate two such modulations and so give insight into whether they are likely to reflect the same or different brain processes. Similarly, if multiple stimulus features or aspects are presented in an experiment, the PID can be applied to quantify how much of the neural response is commonly predicted from both stimulus features, uniquely available from each or synergistically available only from the combination.

The \Id method of~\cite{JEC} is a very general one and it could be applied to systems other than discrete systems~\cite{JEC} or the Gaussian PIDs developed here. For example, the \Id method could be used with  other types of graphical model, such as mixed discrete-continuous systems~\cite{JW, SLL} based on the CG model, and also  in  multivariate autoregressive modelling of time series data~\cite{barrett, PN, KS} using graphical models~\cite{dal, brill, SDV}. We look forward to engaging in further exploration of the potential of the \Id method. 



\begin{appendices}

\setcounter{equation}{0}
\renewcommand\theequation{A.\arabic{equation}}
\section{Proof of Proposition 2}

Examination of the edge values in Table~\ref{edges1} shows that the PID derived here satisfies equations B2-B5 of~\cite{JEC}, on taking $m=0.$ Therefore, the properties of consistency, non-negativity, self-redundancy, monotonicity and identity follow for this new PID using the arguments given in~\cite{JEC}. Therefore, we consider only parts (b)-(h). \newline

\noindent (b) This is true because both $b$ and $d$ are equal to $I(X_0;Y)$.

\noindent (c) When $\text{unq0} =i$, the unique informations and the redundancy components are
$$  \frac{1}{2} \log \frac{1-q^2r^2}{1 - q^2},   \quad \frac{1}{2} \log \frac{1-q^2r^2}{1 - r^2},  \quad \frac{1}{2} \log \frac{1}{1 - q^2 r^2}  $$ respectively, and all these terms are independent of $p$.

\noindent (d) We are given that $q \neq 0, r \neq 0.$  Now,
$$ b - i = \frac{1}{2} \log \frac{1}{1 - q^2 r^2} > 0$$ when $q \neq 0, r \neq 0,$ and since $|q|<1, |r|<1.$ Hence the minimum of the edge values is not $b$ or $d$, which leaves only $i$ and $k$ as possibilities.

\noindent (e) From~$\eqref{condMI1}$ and the expression for $k$ in Table~\ref{edges1}, we see that the synergy component $$I(X_0;Y|X_1) - \text{unq0}$$ is equal to zero.

\noindent (f, g) We will use the definitions of the \Id and \Imm PIDs in~$\eqref{idep1}$-$\eqref{idep4}$ and~$\eqref{mmi1}$-$\eqref{mmi4}$ and also the bivariate and conditional mutual informations in~$\eqref{bivMI0}$-$\eqref{condMI2}$. We denote the \Imm redundancy and synergy components by $R_m$ and $S_m$, respectively, using $R_d$ and $S_d$ for the corresponding \Id components. Denote unq0 and unq1 in \Id as $U_{0d}, U_{1d}$, respectively.  We note that 
\begin{equation} I(X_0;Y) =0 \iff q=0, \quad \text{and} \quad I(X_0;Y|X_1) =0 \iff q = p r.  \label{equal1}
\end{equation} 
First, suppose that $I(X_0;Y) < I(X_1;Y).$ If $I(X_0;Y) =  0$, then $q=0$ and $0=b =i <k$ in \Id. Hence, $$R_m = R_d =0,\quad  \text{and} \quad  S_m = S_d = I(X_0;Y|X_1). $$ If $I(X_0;Y) \neq 0$, then $R_m = I(X_0;Y)$ and $S_m =I(X_0;Y|X_1).$ 
In \Id, the redundancy and synergy components are 
$$ R_d = I(X_0;Y) - U_{0d}, \quad \text{and} \quad S_d = I(X_0;Y|X_1) - U_{0d}. $$ It follows that $R_m \geq R_d$ and $S_m \geq S_d$ with equality iff $U_{0d}=0$.
From (d), it follows for \Id that $U_{0d}= i$ or $U_{0d} =k$.  Since $I(X_0;Y) \neq 0$, $i >0$ so $U_{0d} =0$ iff $k=0$, which from Table~\ref{edges1} and~$\eqref{equal1}$ is true iff $I(X_0;Y|X_1)=0$, in which case $S_m = S_d =0$ and $R_m = R_d =I(X_0;Y).$ Hence result. 

The proof when $I(X_0;Y) >I(X_1;Y)$ is similar and is omitted, although it is worth noting that
\begin{equation} I(X_1;Y) =0 \iff r=0, \quad \text{and} \quad I(X_1;Y|X_0) =0 \iff r = p q. \label{equal2} \end{equation}

When $I(X_0;Y) = I(X_1;Y)$, then
 $$R_m = I(X_0;Y) =I(X_1;Y), \quad \text{and} \quad R_d = I(X_0;Y) - U_{0d} =I(X_1;Y) - U_{1d},$$ 
$$S_m= I(X_0;Y|X_1) =I(X_1;Y|X_0), \,\, \text{and} \,\, S_d = I(X_0;Y|X_1) - U_{0d}=  I(X_0;Y|X_1) - U_{1d}. $$ Therefore 
$R_m \geq R_d$ and $ S_m \geq S_d$ with equality iff $U_{0d} =\ U_{1d} =0$. From the argument above, this happens iff
$$ I(X_0;Y) = I(X_1;Y) =0, \quad \text{or} \quad I(X_0;Y|X_1) = I(X_1;Y|X_0) =0. $$ In this case, the \Id and \Imm redundancy and synergy components are equal if, and only if, each of  $X_0$ and $X_1$ is independent of $Y$, and each of $X_0$ and $X_1$ is conditionally independent of $Y$ given the other predictor.

\noindent (h) When $I(X_0;Y|X_1) =0$, the unq0 and syn   components are zero in both the \Id and \Imm PIDs. From~$\eqref{equal1}$, $ p = q r$, and from Table~\ref{edges1} we see that in the \Id PID, $i =k < b$,  and so \linebreak $\text{U}_{1d} = I(X_1;Y|X_0)$ and $\text{R}_d = I(X_0; Y).$ 
Since $(X_0;Y|X_1) =0$, it follows from 
$$ I(X_0, X_1; Y) = I(X_0;Y) + I(X_0;Y|X_1) = I(X_1;Y) + I(X_1;Y|X_1) $$
that $I(X_0;Y) \leq I(X_1;Y)$ and so in the \Imm PID, $ \text{red} = I(X_0;Y)$ and $ \text{unq1} = I(X_1;Y|X_0).$  
It follows that the \Id and \Imm PIDs are identical.

\noindent The proof when $I(X_1;Y;X_0)=0$ is very similar and it is  omitted. Model $U_6$ has $I(X_0;Y|X_1) =0$, model $U_5$ has $I(X_1;Y|X_0) =0$, and models $U_1 \ldots U_4$ have at least one of these conditions. Hence result.

In model $U_7$, $p=q r$. If $q \neq 0, r \neq 0$, it follows from~$\eqref{cov1}$ and Table~\ref{edges1}   that 
$$ |\Sigma_Z| = (1-q^2) (1-r^2), \quad \text{and} \quad I(X_0;Y|X_1) = k = i  < b\,\,(\text{if}\,\,q \neq 0).$$ Therefore, in the \Id PID unq0 =k and so syn =0. If $q=0$ then from Table~\ref{edges1}, $b=i=k= I(X_0;Y|X_1) =0, $ and so syn =0. If $r=0$, then
$b =i =k =I(X_0;Y|X_1)$, and so syn =0. Hence result.

\section{Proof of Matrix Lemmas}
We begin by stating some some useful results from matrix algebra~\cite[p. 472]{HJ},~\cite[p. 475]{CDM}.

Suppose that a  symmetric matrix $M$ is partitioned as
$$ M = \begin{bmatrix} A & B \\ B^T & C \end{bmatrix}, \label{partM}$$ where $A$ and $C$ are symmetric and square. Then \begin{itemize}
\item[(i)] The matrix $M$ is positive definite if and only if $A$ and $ C - B^T A^{-1}B$ are positive definite.
\item[(ii)]The matrix $M$ is positive definite if and only if $C$ and $ A - B C^{-1}B^T$ are positive definite.
\item[(iii)]$ |M| = |A| |D - B^T C^{-1} B|.$
\end{itemize}

\begin{flushleft}
{\bf Proof of Lemma 1}
\end{flushleft}
If we write $M$ as 
$$ M= \begin{bmatrix} A & M_{13} \\ M_{13}^T & M_{33} \end{bmatrix}, \quad \text{with}  \quad A = \begin{bmatrix} M_{11} & M_{12} \\ M_{12}^T & M_{22} \end{bmatrix} $$
then if $M$ is positive definite it follows from (i) that $A$ and $M_{33}$ are both positive definite. Applying result (i) to the matrix $A$ then shows that $M_{11}$ and $M_{22}$ are also positive definite. A positive definite matrix is nonsingular. Hence result.
\begin{flushleft}
{\bf Proof of Lemma 2}
\end{flushleft}
Given that $\Sigma_Z$ is positive definite, we note that 
$$ \begin{bmatrix} \id{n_0} & P \\ P^T & \id{n_1} \end{bmatrix}, \quad  \begin{bmatrix} \id{n_0} & Q \\ Q^T & \id{n_2} \end{bmatrix}, \quad 
 \begin{bmatrix} \id{n_1} & R \\ R^T & \id{n_2} \end{bmatrix}
$$ are principal sub-matrices of $\Sigma_Z$ and so they are positive definite~\cite[p. 397]{HJ}. From (i, ii), it follows that the matrices
$$ \id{n_1} - P^TP,\,\, \id{n_0} - PP^T, \,\,\id{n_2}-R^TR,\,\, \id{n_1}-RR^T, \,\,\id{n_2}-Q^TQ, \,\,\id{n_0}-QQ^T.$$
are positive definite.

Suppose that $I_n - X^TX$ is positive definite, where $X$ is a $p \times n$ matrix. Then  the matrix $X^TX$ is positive semi-definite and so has non-negative eigenvalues, $\lambda_1, \lambda_2, \ldots \lambda_n$. The eigenvalues of 
$I_n - X^TX$ are 
$\{ 1 - \lambda_i: i =1, 2, ..., n\}$. Since $I_n -X^TX$ is positive definite we know that 
$1 - \lambda_i >0$ for $i=1, 2, ..., n.$ It follows that $0 < 1 - \lambda_i \leq 1$ for $i=1, 2, ..., n.$ 
Since the determinant of a square matrix is the product of its eigenvalues we have that 

$$|I_n -X^TX| = \prod_{i=1}^n (1 - \lambda_i),$$ and so $ 0 < |I_n -X^TX| \leq 1$. It also follows that $|I_n -X^TX| =1$ if, and only if, all the eigenvalues of $X^TX$ are equal to zero, which means that $X^TX$ is the zero matrix.Taking $X=P, P^T, Q, Q^T, R, R^T$ in turn gives the required result.

Application of (iii) gives the result that
$$ \begin{vmatrix}  \id{n_0} & P \\ P^T & \id{n_1}  \end{vmatrix}  = |\id{n_1} -P^TP |. $$

\setcounter{equation}{0}
\renewcommand\theequation{C.\arabic{equation}}
\section{Proof of Proposition 4}
Examination of the edge values in Table~\ref{ev2} shows that the multivariate \Id PID derived here satisfies equations B2-B5 of~\cite{JEC}, on taking $m=0.$ Therefore, the properties of consistency, non-negativity, self-redundancy, monotonicity and identity follow for this new PID using the arguments given in~\cite{JEC}. Hence, we focus attention only on parts (b-h). \newline

\noindent (b) This is true since both $b$ and $d$ are equal to $I(\mathbf{X}_0; Y).$

\noindent (c) When unq0 =$i$, then the expressions for the unique informations and the redundancy, given in Table~\ref{ev2} and~$\eqref{mid1}$-$\eqref{mid2}$ do not depend on the matrix $P$

\noindent (d) From~$\eqref{trans}$  and the assumption that neither $ \Sigma_{02}$ nor $\Sigma_{12}$ is equal to a zero matrix, it follows that $QR^T$ is not equal to a zero matrix.  From Table~\ref{ev2}, 
$$ b - i = \frac{1}{2} \log \frac{1}{|\id{n_1} - RQ^TQR^T|}.$$ From Table 5, we have that the covariance matrix under model $M_7$ is
$$ \Sigma_7 = \begin{bmatrix} \id{n_0} & QR^T & Q\\  RQ^T & \id{n_1}& R\\ Q^T& R^T& \id{n_2} \end{bmatrix}. $$
Applying a similar argument to that in the proof of Lemma 2, it follows that $| \id{n_1} - RQ^TQR^T|$ is positive and bounded above by unity. Also it is equal to unity only when $QR^T$ is equal to the zero matrix having $n_0$ rows and $n_1$ columns. Since this is not the case, it follows that
\begin{equation} 0 < |\id{n_1} -RQ^TQR^T| < 1 \label{QR} \end{equation} and so $ b > i$. Therefore the minimum does not occur at $b$ or $d$, leaving only $i$ and $k$ as the remaining possibilities.

\noindent (e) From~$\eqref{mcondMI1}$ and the entry for $k$ in Table~\ref{ev2}, we see that the synergy component is equal to zero when $\text{unq0}=k$.

\noindent (f, g) The proofs are very similar to those for (f, g) in Proposition 2 and so they are omitted. The following results are useful.
From~$\eqref{mvMI1}$-$\eqref{mvMI4}$, we can state the following results.

$I(\mathbf{X}_0; \mathbf{Y})=0$\W\W\W iff $\mathbf{X}_0$ and $\mathbf{Y}$ are independent, iff the matrix $P$ is a zero matrix.  

$I(\mathbf{X}_1; \mathbf{Y})=0$\W\W\W  iff $\mathbf{X}_1$ and $\mathbf{Y}$ are independent, iff the  matrix $R$ is a zero matrix.

 $I(\mathbf{X}_0; \mathbf{Y}| \mathbf{X}_1) =0$ iff $\mathbf{X}_0$ and $\mathbf{Y}$ are conditionally independent given $\mathbf{X}_1$, iff $Q =P R$, from~$\eqref{v}$. 
 
 $I(\mathbf{X}_1; \mathbf{Y}| \mathbf{X}_0) =0$ iff $\mathbf{X}_1$ and $\mathbf{Y}$ are conditionally independent given $\mathbf{X}_0$, iff $R =P^TQ$, from~$\eqref{w}$.

\noindent (h) When $I(\mathbf{X}_0;\mathbf{Y}|\mathbf{X}_1) =0$, the unq0 and syn  components components are zero in both the \Id and \Imm PIDs, and so  in the \Id PID, $\text{unq1} = I(\mathbf{X}_1;\mathbf{Y}|\mathbf{X}_0)$ and $\text{red} = I(\mathbf{X}_0; \mathbf{Y}).$ 
Since $I(\mathbf{X}_0;\mathbf{Y}|\mathbf{X}_1) =0$, it follows that 
$I(\mathbf{X}_0; \mathbf{Y}) \leq I(\mathbf{X}_1;\mathbf{Y})$ and so in the \Imm PID, $ \text{red} = I(\mathbf{X}_0; \mathbf{Y})$ and $ \text{unq1} =  I(\mathbf{X}_1;\mathbf{Y}|\mathbf{X}_0).$  
It follows that the \Id and \Imm PIDs are identical.

\noindent The proof when $I(\mathbf{X}_1;\mathbf{Y};\mathbf{X}_0)=0$ is very similar and it is  omitted. Model $M_6$ has $I(\mathbf{X}_0;\mathbf{Y}|\mathbf{X}_1) =0$, model $M_5$ has $I(\mathbf{X}_1;\mathbf{Y}|\mathbf{X}_0) =0$, and models $M_1 \ldots M_4$ have at least one of these conditions. Hence result.

In model $M_7$, $P=QR^T, $ from Table~\ref{CIcon2}.  Also from~$\eqref{trans}$
$$ QR^T =  \Sigma_{00}^{-\tfrac{1}{2}} \Sigma_{02} \Sigma_{22}^{-1}  \Sigma_{12}^T \Sigma_{11}^{-\tfrac{1}{2}}. $$ Provided that neither $ \Sigma_{02}$ nor $\Sigma_{12}$ is equal to a zero matrix,  it follows from~$\eqref{sig2}$, Table~\ref{mitot2}  and~$\eqref{QR}$ that 
$$ |\Sigma_Z| = |\id{n_2} -Q^TQ| |\id{n_2} - R^TR|, \quad \text{and} \quad I(X_0;Y|X_1) = k = i  < b.$$   Therefore, in the \Id PID, unq0 =k and so syn =0. If $ \Sigma_{02}$  is equal to a zero matrix, then from Table~\ref{ev2}, $b=i=k= I(X_0;Y|X_1) =0, $ and so syn = 0.   Similarly, if $\Sigma_{12}$ is equal to a zero matrix then $b=i=k= I(X_0;Y|X_1)$, and so syn =0.  Hence result.

\setcounter{equation}{0}
\renewcommand\theequation{D.\arabic{equation}}
\section{Computation of the Multivariate \Id PID}
Given a multivariate data having two different sets of predictors, $\mathbf{X}_0, \mathbf{X}_1$ and a target, $\mathbf{Y}$, the special formulae presented in Table~\ref{ev2} and~$\eqref{mid1}$-$\eqref{mid2}$ can be used to compute the \Id PID. In order to ensure that the input data have the required format one can use the following procedure.

Suppose that the general covariance matrix is

\begin{equation}\Sigma = \begin{bmatrix}  \Sigma_{00} & \Sigma_{01} & \Sigma_{02}\\ \Sigma_{01}^T & \Sigma_{11} & \Sigma_{12}\\ \Sigma_{02}^T & \Sigma_{12}^T & \Sigma_{22}
  \end{bmatrix}. \label{gencov} \end{equation}

\noindent Then the required matrices, $P, Q, R,$ can be obtained from this covariance matrix using the following formulae based on~$\eqref{trans}$. The transposes are used here since the extracted square root matrix used here is not symmetric.

\begin{equation} P = \left[ \Sigma_{00}^{-\tfrac{1}{2}} \right]^T \Sigma_{01} \Sigma_{11}^{-\tfrac{1}{2}}, \quad Q = \left[\Sigma_{00}^{-\tfrac{1}{2}} \right]^T\Sigma_{02} \Sigma_{22}^{-\tfrac{1}{2}}, \quad R = \left[ \Sigma_{11}^{-\tfrac{1}{2}} \right]^T \Sigma_{12} \Sigma_{22}^{-\tfrac{1}{2}}. \label{transD} \end{equation}
Therefore, the procedure involves: (a) extracting the block diagonal matrices, $\Sigma_{00}, \Sigma_{11}, \Sigma_{22},$ \linebreak in~$\eqref{gencov}$, (b) finding each square root as an upper triangular matrix by Cholesky decomposition, (c) inverting the upper triangular matrix  using a 'backsolve' method and (d) applying the formulae in~$\eqref{transD}.$ Code written in R was used here and in the other examples.

\section{Deviance tests}
We give some details of the deviance tests that have been performed in Sections 2.6 and 3.6. The following notes are based on~\cite[p. 40]{DE}. See also~\cite{JW}. Suppose that a random vector $\mathbf{Z}, $  of dimension $q$, follows a Gaussian  graphical model $\cal{M}$ having  mean vector  $\boldsymbol{\mu}$ and covariance matrix $\Sigma$. Suppose that a sample of $N$ observations is available and that the sample covariance matrix (with divisor $N$) is $S$. Let the estimated covariance matrix for model $\cal{M}$ be $\hat{\Sigma}$. Then the maximised log likelihood under model $\mathcal{M}$ is
$$ \hat{l}_m = - N q \ln(2 \pi)/2 - N \ln|\hat{\Sigma}|/2 - N q /2. $$ Under the full (or saturated)  model $\mathcal{M}_{f}$, $\hat{\Sigma} =S$ and so the maximised likelihood under $\mathcal{M}_f$ is
$$ \hat{l}_f = - N q \ln(2 \pi)/2 - N \ln|S|/2 - N q /2. $$
The deviance of a model is defined to be 
$$ G^2 = 2 ( \hat{l}_f - \hat{l}_m) = N \ln \frac{|\hat{\Sigma}|}{|S|} $$ and $G^2$ can be used as a test statistic when testing model $\cal{M}$ within the saturated model $\mathcal{M}_f$. The null distribution of $G^2$ has an asymptotic chi-squared distribution with degrees of freedom given by the difference in the number of edges between model $\cal{M}$ and the saturated model, $\mathcal{M}_f$. This test is an example of a generalised likelihood ratio test which can be used to compare nested statistical models. A $p$ value can be calculated as 
$p = \Pr(G^2 \geq G^2_{\text{obs}} |\,\, \cal{M} \,\, \text{is true}), $ where $G^2_{\text{obs}}$ is the observed value of the test statistic $G^2$. When $p < 0.01$ we may say that there is strong evidence against model $\cal{M}$, the implication being that this model does not provide an acceptable fit to the data. On the other if $p >0.1$ we may say that there is little evidence against model $\cal{M}$, with the implication being that this model provides an acceptable fit to the data. We may say that there is moderate evidence if $0.01 < p < 0.05,$ and weak evidence if $0.05 < p < 0.1$  in the  borderline case. 

It should be noted that this test is approximate and its performance improves the larger the sample is. There are exact tests in some cases and there are correction factors that can improve the approximation~\cite{ JW, DE}. These were not used in this study because the results are so clear cut.

When model $\mathcal{M}_0$ is a special case of (or nested within) model $\mathcal{M}_1$ and it is required to test model $\mathcal{M}_0$ within model $\mathcal{M}_1$ then the deviance test statistic is 
$$ D = N \ln \frac{|\hat{\Sigma}_0|}{|\hat{\Sigma}_1|}$$
and the null distribution of $D$ also has an asymptotic chi-squared distribution with degrees of freedom equal to the difference in the number of edges between $\mathcal{M}_0$ and $\mathcal{M}_1$. 

There is an interesting connection between the test statistic $D$ and some of the edge values in Fig.~\ref{fig1}. For each of the edge values in the set $\{ b, c, d, f, j, k\}$, the edge value is equal to the corresponding value of the test statistic $D$ divided by $2 N \ln 2$. This is not the case for edge values $h$ and $i$.

\end{appendices}

\end{document}